\documentclass[fleqn,usenatbib]{mnras}

\usepackage{newtxtext,newtxmath}

\usepackage[T1]{fontenc}

\DeclareRobustCommand{\VAN}[3]{#2}
\let\VANthebibliography\thebibliography
\def\thebibliography{\DeclareRobustCommand{\VAN}[3]{##3}\VANthebibliography}

\usepackage{graphicx}	
\usepackage{amsmath}	

\newcommand{\fesc}{$f_{\mathrm{esc}}$}

\title[Testing an indirect method for probing LyC leakage]{Testing an indirect method for identifying galaxies with high levels of Lyman continuum leakage}

\author[S. Yamanaka et al.]{
Satoshi Yamanaka,$^{1,2}$\thanks{E-mail: syamanaka@aoni.waseda.jp}
Akio K. Inoue,$^{1,2,3}$
Toru Yamada,$^{4}$
Erik Zackrisson,$^{5}$
\newauthor
Ikuru Iwata,$^{6}$
Genoveva Micheva,$^{7}$
Ken Mawatari,$^{8}$
Takuya Hashimoto$^{1,2,6,9}$
\newauthor
and Mariko Kubo$^{6}$
\\
$^{1}$Waseda Research Institute for Science and Engineering, Faculty of Science and Engineering, Waseda University,\\
3-4-1, Okubo, Shinjuku, Tokyo 169-8555, Japan\\
$^{2}$Department of Environmental Science and Technology, Faculty of Design Technology, Osaka Sangyo University,\\
3-1-1, Nakagaito, Daito, Osaka 574-8530, Japan\\
$^{3}$Department of physics, School of Advanced Science and Engineering, Faculty of Science and Engineering, Waseda University,\\
3-4-1, Okubo, Shinjuku, Tokyo 169-8555, Japan\\
$^{4}$Institute of Space and Astronautical Science, Japan Aerospace Exploration Agency, 3-1-1, Yoshinodai, Chuo-ku, Sagamihara, Kanagawa 252-5210, Japan\\
$^{5}$Observational Astrophysics, Department of Physics and Astronomy, Uppsala University, Box 516, SE-751 20 Uppsala, Sweden\\
$^{6}$National Astronomical Observatory of Japan, 2-21-1 Osawa, Mitaka, Tokyo 181-8588, Japan\\
$^{7}$Leibniz-Institut f\"ur Astrophysik, An der Sternwarte 16, D-14482 Potsdam, Germany\\
$^{8}$Institute for Cosmic Ray Research, The University of Tokyo, 5-1-5 Kashiwa-no-Ha, Kashiwa, Chiba 277-8582, Japan\\
$^{9}$Tomonaga Center for the History of the Universe (TCHoU), Faculty of Pure and Applied Sciences, University of Tsukuba, Tsukuba, Ibaraki 305-8571, Japan
}

\date{Accepted XXX. Received YYY; in original form ZZZ}

\pubyear{2020}

\begin{document}
\label{firstpage}
\pagerange{\pageref{firstpage}--\pageref{lastpage}}
\maketitle

\begin{abstract}
 Using a sample of galaxies at $z\approx 3$ with detected Lyman Continuum (LyC) leakage in the SSA22 field, we attempt to verify a proposed indirect method for identifying cases with high LyC escape fraction \fesc\ based on measurements of the H$\beta$ equivalent width (EW) and the $\beta$ slope of the UV continuum. To this end, we present Keck/MOSFIRE H$\beta$ emission line flux measurements of LyC galaxies (LCGs) at spectroscopic redshifts $z_{\mathrm{spec}} \sim 3.3$, Lyman break galaxies (LBGs) at photometric redshifts $z_{\mathrm{phot}} = 2.7$--$3.7$, and Ly$\alpha$ emitters at $z_{\mathrm{phot}} = 3.1$. We also reconfirm the spectroscopic redshifts and measure the H$\beta$ emission line fluxes from 2 LCGs and 6 LBGs. For the LCG in our sample with the most extreme \fesc, as revealed by the direct detection of LyC photons, we find that the EW(H$\beta$)--$\beta$ method gives a broadly consistent estimate for \fesc, although the error bars remain very large. We also discuss how a combination of $f_\mathrm{esc}$ measurements based on direct and indirect methods can shed light on the LyC escape mechanism and the anisotropy of the leakage.
\end{abstract}




\section{Introduction}

Cosmic reionization is a phase transition from the neutral to ionized state of the intergalactic medium (IGM), likely driven by ionizing radiation from star-forming galaxies (SFGs) and/or active galactic nuclei \citep[AGNs; e.g.][]{Mad99ApJ...514..648M, Mad15ApJ...813L...8M, Rob15ApJ...802L..19R}. Various observations indicate that the reionization process was completed by $z \sim 6$: the Gunn--Peterson optical depth \citep[e.g.][]{Fan02AJ....123.1247F, Fan06AJ....132..117F}, the clustering of Ly$\alpha$ Emitters \citep[LAEs; e.g.][]{Ouc10ApJ...723..869O, Ouc18PASJ...70S..13O}, the evolution of the luminosity function of LAEs \cite[e.g.][]{Ksk06ApJ...648....7K, Ksk11ApJ...734..119K, Ito18ApJ...867...46I, Kon18PASJ...70S..16K}, the fraction of Ly$\alpha$ emitting galaxies in SFGs \citep[e.g.][]{Ono12ApJ...744...83O, Sch12ApJ...744..179S}, the Ly$\alpha$ damping wing in gamma-ray burst after-glow spectra \citep[e.g.][]{Tot06PASJ...58..485T, Tot16PASJ...68...15T, Gre09ApJ...693.1610G}, and the kinetic Sunyaev--Zeldovich effect \citep[e.g.][]{Zah12ApJ...756...65Z, Pla16A&A...596A.108P}. However, despite a great deal of effort to understand cosmic reionization, the detailed history, sources, and topology of this epoch are not yet understood.

In order to understand the dominant sources of cosmic reionization, the fraction of ionizing photons (Lyman continuum, hereafter LyC, at $\lambda_{\mathrm{rest}} < 912$\AA) that escapes from SFGs/AGNs into the surrounding IGM, \fesc, is one of the most important physical quantities. For reionization by SFGs, a LyC escape fraction on the order of 10\%  seems to be required \citep[e.g.][]{Ino06MNRAS.371L...1I, Fin15ApJ...810...71F, Fin19ApJ...879...36F, Bou16ApJ...831..176B}. Based on current constraints, the contribution from AGNs to the reionization process moreover appears to be subdominant compared to that of SFGs \citep[e.g.][]{Mic17MNRAS.465..302M, Mat18ApJ...869..150M, Kul19MNRAS.488.1035K}.

The simplest and most robust way to constrain \fesc\ is direct imaging and/or direct spectroscopic observations of the escaping LyC photons. The standard procedure is to estimate \fesc\ of SFGs/AGNs from their observed luminosity ratio of ionizing to non-ionizing ultraviolet (UV) radiation, $L_{\mathrm{UV,obs}}/L_{\mathrm{LyC,obs}}$, using assumptions on the intrinsic luminosity ratio of $L_{\mathrm{UV,int}}/L_{\mathrm{LyC,int}}$, the IGM absorption along its sightline, and the dust attenuation \citep[e.g.][]{Ste01ApJ...546..665S, Ino05A&A...435..471I, Sia07ApJ...668...62S}. 

Since it is next to impossible to directly observe the LyC photons escaping from SFGs/AGNs at $z > 5$ due to foreground IGM absorption \citep{Ino08MNRAS.387.1681I}, the LyC observations for lower-$z$ analogs are important for inferring the likely \fesc\ of SFGs/AGNs in the epoch of cosmic reionization, and to test indirect methods of estimating $f_\mathrm{esc}$.

Numerous attempts have been made to estimate the LyC escape fractions of individual SFGs at $z < 4$. Since Earth's atmosphere is very efficient in absorbing far-UV radiation, space telescopes such as the Far Ultraviolet Spectroscopic Explorer (\textit{FUSE}) and the Hubble Space Telescope (\textit{HST}) have very been important for studying the \fesc\ of local and/or low-$z$ SFGs. Several detections of direct LyC radiation from $z < 1$ starburst galaxies have been made, typically with $f_{\mathrm{esc}} \lesssim 10$\% \citep[e.g.][]{Brg06A&A...448..513B, Lei11A&A...532A.107L, Lei13A&A...553A.106L, Bor14Sci...346..216B, Izo16MNRAS.461.3683I, Izo16Natur.529..178I}, although some objects have been reported with $f_{\mathrm{esc}} \sim 40$--$70$\% \citep{Izo18MNRAS.474.4514I, Izo18MNRAS.478.4851I}. At $z \approx 2$--$4$, the \fesc\ values of SFGs have been investigated by both space and ground based telescopes. In this redshift range, many individual objects appear to show relatively high \fesc\ ($f_{\mathrm{esc}} \gtrsim 10$\%) compared to the local value \citep[e.g.][]{Ste01ApJ...546..665S, Ste18ApJ...869..123S, Iwa09ApJ...692.1287I, Iwa19MNRAS.488.5671I, Mos15ApJ...810..107M, Sha16ApJ...826L..24S, Van16ApJ...825...41V, Van18MNRAS.476L..15V, Bia17ApJ...837L..12B, Mic17MNRAS.465..316M, Nai17ApJ...847...12N, Fle19ApJ...878...87F}.

At the same time, attempts to estimate the {\it typical} \fesc\ using stacked samples at $z = 1$--$3$ indicate that this value may be much lower. These stacking analyses often result in non-detections of LyC, with corresponding upper limits at $f_{\mathrm{esc}} <$ several \% \citep[e.g.][]{Sia07ApJ...668...62S, Sia10ApJ...723..241S, Van10ApJ...725.1011V, Gra16A&A...585A..48G, Mic17MNRAS.465..316M}. A very low typical $f_{\mathrm{esc}} \approx 0.5$ \% has also been derived from the study of gamma-ray bursts at $z \approx 1.6$--$6.7$ \citep{Tanvir19}. Hence, the average \fesc\ at these intermediate redshifts may be on the low side of what would be required for cosmic reionization \citep[but see][for an attempt to explain reionization with $f_{\mathrm{esc}} \leq 5$ \%]{Fin19ApJ...879...36F}. Whether the average \fesc\ increases as one approaches the epoch of reionization ($z>6$) remains an open question. 

Since studies of individual objects with LyC detections report  significant \fesc\ values (from several \% at $z \sim 0$ to $10$--$60$\% at $z \sim 3$--$4$), it is likely that the LyC escape depends strongly on physical quantities such as the properties of gas and dust in the ISM, the viewing angle, the star formation rate, and/or the stellar mass \citep[e.g.][]{Nes11ApJ...736...18N, Nes13ApJ...765...47N, Mos13ApJ...779...65M}. For a statistical discussion, it is important to collect data on LyC-detected SFGs which cover a wide parameter space in terms of physical characteristics. In numerical simulations, less-massive and/or UV fainter SFGs, which are more abundant than massive and/or UV brighter SFGs, are for instance often predicted to have high \fesc\ and hence contribute significantly to cosmic reionization \citep[e.g.][]{Yaj11MNRAS.412..411Y, Wis14MNRAS.442.2560W, Paa15MNRAS.451.2544P}.

Depending on the physical mechanism behind the leakage, this can have number of distinct effects on various spectral features. When the escape path is optically thin to Ly$\alpha$ photons ($\tau_{\mathrm{Ly}\alpha} \sim 1$ for $\log N_{\mathrm{HI}} \sim 13$), the escape path is also optically thin to LyC photons ($\tau_{\mathrm{LyC}} \sim 1$ for $\log N_{\mathrm{HI}} \sim 17$). Hence, one expects a positive correlation between \fesc\ and the emission-line equivalent width (EW) of Ly$\alpha$ \citep[e.g.][]{Mic17MNRAS.465..316M, Ste18ApJ...869..123S}. When assuming the link of the escape path between LyC and Ly$\alpha$, the line profile of Ly$\alpha$ also becomes a probe for \fesc\ \citep{Ver15A&A...578A...7V, Ver17A&A...597A..13V, Dij16ApJ...828...71D}. The luminosity ratio of [\ion{O}{iii}] $\lambda\lambda\, 4959, 5007$ and [\ion{O}{ii}] $\lambda 3727$, [\ion{O}{iii}]/[\ion{O}{ii}], has also been suggested as a probe for high-\fesc\ SFGs \citep{Nak13ApJ...769....3N, Nak14MNRAS.442..900N}. Indeed, very high \fesc\ ($f_{\mathrm{esc}} \gtrsim 50$\%) are found among SFGs with high [\ion{O}{iii}]/[\ion{O}{ii}] at both low and high $z$ \citep{deB16A&A...585A..51D, Van16ApJ...825...41V, Van18MNRAS.476L..15V, Izo18MNRAS.474.4514I}. Even so, \citet{Nak19arXiv190907396N} report that no LyC escape is detected for some SFGs despite a high [\ion{O}{iii}]/[\ion{O}{ii}] ratio. Hence, it seems that a high [\ion{O}{iii}]/[\ion{O}{ii}] ratio may be a necessary, rather than sufficient condition for LyC leakage.

Although there have been attempts to determine \fesc\ for SFGs at $z < 4$ from direct LyC measurements, our knowledge of \fesc\ is still far from complete. There are essentially five difficulties in estimating \fesc\ from direct LyC measurements.
First of all, it is hard to directly observe the escaping LyC photons from high-$z$ SFGs (and almost impossible from SFGs at $z > 5$) due to the foreground absorption by \ion{H}{i} clouds in the IGM such as Lyman limit systems \citep[e.g.][]{Ino08MNRAS.387.1681I, Ino14MNRAS.442.1805I}. Second, significant sightline-to-sightline variations in IGM absorption are predicted \citep{Ino08MNRAS.387.1681I,Vasei16,Ste18ApJ...869..123S}, which may be difficult to assess for individual targets. In fact, \citet{Riv19Sci...366..738R} report on large variations among the multiply-imaged LyC spots from a gravitationally lensed SFG at $z = 2.4$. Third, it is possible that the observed flux, which is assumed to be due to escaping LyC photons, does not come from SFGs at high $z$ but from the foreground (low-$z$) interlopers \citep{Van10MNRAS.404.1672V, Sia15ApJ...804...17S}. Fourth, the leakage may be anisotropic, so that the $f_\mathrm{esc}$ measured in our direction may deviate significantly from the total $f_\mathrm{esc}$ of the target galaxy. Finally, because the UV light especially at $\lambda_{\mathrm{obs}} \lesssim 3000$\AA\ is absorbed by the Earth's atmosphere, we need space telescopes for the direct LyC measurements of low-$z$ objects. In order to investigate the redshift evolution of \fesc\ and the detailed correlations between \fesc\ and other physical quantities, we would greatly benefit from an indirect method which can avoid these problems.

\citet{Zac13ApJ...777...39Z} have proposed a scheme to indirectly assess \fesc\ from two observational quantities not affected by IGM attenuation -- the H$\beta$ line equivalent width (EW) and the rest-frame intrinsic UV spectral slope $\beta$, defined as $f_{\lambda} \propto \lambda^{\beta}$ at $\lambda_{\mathrm{rest}} \sim$ 1300--2500\AA\ \citep{Cal94ApJ...429..582C}. This EW(H$\beta$)--$\beta$ method is based on the simple idea that a high fraction of LyC photons absorbed in \ion{H}{ii} regions of SFGs should results in stronger nebular emission and a larger EW(H$\beta$). The equivalent width of nebular emission also depends on the production rate of LyC photons, but the UV spectral slope $\beta$ can potentially be used to gauge this, as a blue UV slope is expected to go hand in hand with a high production rate. As a consequence, the distribution of SFGs across the EW(H$\beta$)--$\beta$ diagram could make it possible to single out  high-\fesc\ cases. 

This method was designed for identifying extreme cases of LyC leakage in the $z>6$ galaxy population, which are out of reach of direct LyC detection methods, and originally based on simple toy models for such galaxies. However, \citet{Zac17ApJ...836...78Z} also show the validity of this idea for SFGs at $z > 6$ in cosmological simulations. If verified, this could be a useful tool for studying LyC leakage with the James Webb Space Telescope (\textit{JWST}), which will be able to spectroscopically detect H$\beta$ in large samples of galaxies up to $z \approx 9$.

However, before attempting to apply the EW(H$\beta$)--$\beta$ method to SFGs at $z > 6$, it would be very useful to verify this method with SFGs for which the LyC has been measured through direct means, so that the $f_\mathrm{esc}$ derived from the direct and indirect methods can be compared. This requires applying the method to galaxies at $z<4$, and comes with a number of challenges. Observations have revealed that there is a trend in $\beta$ with redshift in the sense that reionization-epoch galaxies tend to have bluer UV slopes \citep{Bou14ApJ...793..115B}, likely due to evolution in both dust attenuation and stellar population age \citep{Wilkins13}. Without recalibrating the method to simulations at $z<6$, it is therefore necessary to identify a suitable lower-redshift analog with properties that match the $z>6$ galaxy population as closely as possibly. The ideal case would be to detect high levels of LyC from a galaxy with a very blue UV slope, indicative of low dust reddening, although the exact $\beta$ limit for this depends on EW(H$\beta$) as well. The comparison may also be compromised due to the way anisotropic leakage and uncertainties in the IGM line-of-sight absorption may affect the direct $f_\mathrm{esc}$ measurement. Despite these difficulties, we here set out to attempt this test, starting from a sample of $z\approx 3$ galaxies in the SSA22 field\footnote{SSA22 stands for Small Selected Area at R.A. 22h \citep{Cow88ApJ...332L..29C, Lil91ApJ...369...79L}} where a prominent density peak of LBGs, LAEs, Lya blobs and sub-mm galaxies at z=3.09 have been reported \citep{Ste98ApJ...492..428S, Ste00ApJ...532..170S, Hay04AJ....128.2073H, Maty04AJ....128..569M, Tam09Natur.459...61T, Yam12AJ....143...79Y, Ume14MNRAS.440.3462U, Kub15ApJ...799...38K}.

In this paper, we present results of our \textit{K}-band spectroscopic measurements of the H$\beta$ emission line flux from Lyman Continuum Galaxies (hereafter LCGs), which are SFGs with direct LyC detections at $z \sim 3.3$, as well as Lyman break galaxies (LBGs) and LAEs in the SSA22 field. In the end, we are able to identify one very blue LCG for which the EW(H$\beta$)-$\beta$ method may be applied without modifications, and find that the \fesc\ estimated from this technique is broadly consistent with \fesc\ derived from the direct LyC measurement. Stronger constraints (and hence a more decisive test) would, however, require a significant reduction of the observational errors on both EW(H$\beta$) and $\beta$.

In section 2, we describe the details of our sample of LCGs/LBGs/LAEs. In section 3, we describe the spectroscopic observations and the data reductions. In section 4, we explain the method used for measuring the UV spectral slope $\beta$, the emission line flux, and EW(H$\beta$). In section 5, we show our main result, i.e., the EW(H$\beta$)--$\beta$ diagram. In section 6, we discuss the EW(H$\beta$)--$\beta$ method in comparison to the direct LyC measurement. Our results are summarized in section 7. In regard to the cosmological parameters, we assume $\Omega_{m,0} = 0.3$, $\Omega_{\Lambda,0} = 0.7$, $H_{0} = 70\, \mathrm{km\, s^{-1}\, Mpc^{-1}}$. Throughout this work, we adopt the AB magnitude system \citep{Oke83ApJ...266..713O, Fuk96AJ....111.1748F}.

\section{Sample} \label{S2sap}

\subsection{Photometric catalog} \label{S2s1:pc}

\begin{table*}
    \centering
	\caption{Summary of a part of the imaging data complied by Mawatari et al. in preparation.}
	\label{tab1}
    \begin{tabular}{llcccc}
        \hline
        Instrument & Filter & PSF FWHM & Limiting Mag. & $\lambda_{\mathrm{rest, eff}}$ & Reference \\
        & & (original, arcsec) & (smoothed, $2.2^{\prime \prime}\phi$, 5$\sigma$) & (if $z_{\mathrm{source}} = 3.3$) & \\ 
        \hline
        Subaru/Scam & \textit{R}  & 1.08 & 26.5 & 1520 & \citet{Hay04AJ....128.2073H} \\
        & \textit{i}$^{\prime}$ & 0.76 & 26.3 & 1790 &  \citet{Nkm11MNRAS.412.2579N} \\
        & \textit{z}$^{\prime}$ & 0.76 & 25.6 & 2110 & \citet{Nkm11MNRAS.412.2579N} \\
        & \textit{NB359} & 0.84 & 26.1 & 830 &  \citet{Iwa09ApJ...692.1287I} \\
        Subaru/HSC & \textit{y} & 0.56 & 24.1 & 2270 & \citet{Aih18PASJ...70S...8A} \\
        UKIRT/WFCAM & \textit{K} & 0.86 & 22.9 & 5150 &  \citet{Law07MNRAS.379.1599L} \\
        \hline
        \multicolumn{6}{l}{Notes. -- We only list the broad-band and narrow-band filters used for our analysis, i.e., the estimation of UV spectral}\\
        \multicolumn{6}{l}{slope $\beta$ and the measurements of equivaelnt width of H$\beta$.}\\
    \end{tabular}
\end{table*}

In this work, we use the SSA22 \ion{H}{i} Tomography (SSA22HIT) master catalog (Mawatari et al. in preparation). The purpose of the SSA22HIT project is to reveal the spatial distribution of \ion{H}{i} gas in the SSA22 proto-cluster field through the use of Ly$\alpha$ forest tomography \citep{LKG14ApJ...795L..12L, LKG18ApJS..237...31L}. For this purpose, they compile the photometry of available broad-band and narrow-band filters in the SSA22 field, and make a catalog of LAEs and LBGs in a range of photometric redshift ($z_{\mathrm{phot}}$) $= 2.7$--$3.7$. The catalog includes the photometry of broad-band and narrow-band filters of the Canada-France-Hawaii Telescope(CFHT)/MegaCam \citep{Bou03SPIE.4841...72B}, Subaru/Suprime-Cam \citep[Scam;][]{Miy02PASJ...54..833M}, Subaru/Hyper Suprime-Cam \citep[HSC;][]{Miy12SPIE.8446E..0ZM, Miy18PASJ...70S...1M}, Subaru/MOIRCS \citep{Suz03SPIE.4841..307S, Ich06SPIE.6269E..16I}, UKIRT/WFCAM \citep{Cas07A&A...467..777C}, and Spitzer/IRAC \citep{Faz04ApJS..154...10F}. The source detection on the Subaru/Scam \textit{i}$^{\prime}$-band image \citep{Nkm11MNRAS.412.2579N} and multi-band photometry are performed by using SExtractor\footnote{\url{http://www.astromatic.net/software/sextractor}} ver 2.5.0 \citep{Ber96A&AS..117..393B}. Except for the Spitzer data, the point spread function (PSF)-matched images are created by convolving the original images with Gaussian kernels to match a PSF with full width at half maximum (FWHM) $= 1.1^{\prime \prime}$. The PSF FWHM of $1.1^{\prime \prime}$ is adopted because the PSF FWHM of Subaru/\textit{R}-band is $\sim 1.1^{\prime\prime}$, which is larger than that of the other filters except for the Spitzer data. The $2.2^{\prime \prime}$-diameter ($= 2 \times$PSF) aperture flux is measured for the PSF-matched images.

In our analysis, we use the photometry measured for the PSF-matched images of Subaru/Scam \textit{R} \citep{Hay04AJ....128.2073H}, Subaru/Scam \textit{i$^{\prime}$} and \textit{z$^{\prime}$} \citep{Nkm11MNRAS.412.2579N}, Subaru/Scam \textit{NB359} \citep{Iwa09ApJ...692.1287I}, Subaru/HSC \textit{y} \citep[HSC Subaru Strategic Program Public Data Release 1;][]{Aih18PASJ...70S...8A}, and UKIRT/WFCAM \textit{K} \citep[UKIRT Infrared Deep Sky Survey Data Release 10;][]{Law07MNRAS.379.1599L}.
The Subaru/\textit{NB359} is a unique narrow-band filter which directly traces LyC photons from galaxies at $z \gtrsim 3.06$ \citep{Iwa09ApJ...692.1287I, Mic17MNRAS.465..316M}. The central wavelength, the FWHM, and the transmittance larger than $10\%$ are 359nm, 15nm, and [350nm, 371nm], respectively.
We list the detail of the imaging data used in this work in Table \ref{tab1}.

\subsection{Spectroscopic targets} \label{S2s2:st}

 Our spectroscopic sample consists of 42 galaxies in the SSA22 field. Two of them are LCGs at spectroscopic redshifts ($z_{\mathrm{spec}}$) $\sim 3.3$ and the main targets for our MOSFIRE observation. They are selected from a sample of the LyC sources reported by \citet{Iwa09ApJ...692.1287I} and \citet{Mic17MNRAS.465..316M}. In this paper, we will refer to them as LCG-1 and LCG-2. In order to observe these main targets, we set two mask fields, Mask-1 and Mask-2. 37 objects are complementary LBGs at $z_{\mathrm{phot}} \sim 3.2 \pm 0.3$, which are selected from the SSA22HIT master catalog. We will refer to these as SSA22-LBGs in this paper. The remaining three objects are LAEs at $z_{\mathrm{phot}} = 3.1$ \citep{Yam12AJ....143...79Y, Yam12ApJ...751...29Y} which are observed as filler objects (hereafter Y12LAEs). Y12LAE-2 and Y12LAE-3 are taken from the LAE candidates studied in \citet{Yam12AJ....143...79Y} which show significant narrow-band excess but below their criteria for the robust LAE sample. Our sample is summarized in Table \ref{tab2}. The details are described in the following.

\begin{table*}
    \centering
	\caption{Summary of the Mask-1 and Mask-2 samples. }
	\label{tab2}
    \begin{tabular}{rcccccccc}
        \hline
        Name & R.A. & Dec. & $z_{\mathrm{spec}}$ $^{a}$ & \textit{NB359} $^{b}$ & \textit{R} $^{b}$ & \textit{K} $^{b}$ & Reference $^{c}$ \\ 
        \hline
        \multicolumn{8}{c}{Mask-1}\\
        \hline
        LCG-1$^{d}$ & 22:17:08.12 & 00:09:58.08 & 3.287 & 25.18 & 24.69 & 23.46 & (1), (2) \\
        SSA22-LBG-01 & 22:17:13.56 & 00:07:28.93 & 3.087 & 99.00 & 25.17 & 24.19 & (3) \\
        SSA22-LBG-02 & 22:17:06.26 & 00:07:39.03 &  & 99.00 & 25.39 & 99.00 & (3) \\
        SSA22-LBG-03 & 22:17:09.93 & 00:06:02.30 &  & 99.00 & 25.49 & 24.27 & (3) \\
        SSA22-LBG-04 & 22:17:10.55 & 00:05:01.51 &  & 25.71 & 24.99 & 99.00 & (3) \\
        SSA22-LBG-05 & 22:17:04.73 & 00:06:20.91 & 3.112 & 99.00 & 25.59 & 24.55 & (3) \\
        SSA22-LBG-06 & 22:17:04.96 & 00:06:08.30 &  & 99.00 & 25.31 & 99.00 & (3) \\
        SSA22-LBG-07 & 22:17:04.38 & 00:09:20.13 &  & 24.70 & 24.49 & 23.79 & (3) \\
        SSA22-LBG-08 & 22:17:07.97 & 00:05:04.96 &  & 26.51 & 25.40 & 24.18 & (3) \\
        SSA22-LBG-09 & 22:17:15.33 & 00:04:44.39 &  & 26.96 & 24.88 & 99.00 & (3) \\
        SSA22-LBG-10 & 22:17:13.39 & 00:07:52.39 &  & 99.00 & 25.10 & 23.61 & (3) \\
        SSA22-LBG-11 & 22:17:09.94 & 00:07:19.94 &  & 99.00 & 25.29 & 24.14 & (3) \\
        SSA22-LBG-12 & 22:17:13.60 & 00:06:40.05 &  & 27.02 & 24.42 & 23.56 & (3) \\
        SSA22-LBG-13 & 22:17:08.65 & 00:06:03.99 & 2.869 & 26.51 & 24.87 & 23.92 & (3) \\
        SSA22-LBG-14 & 22:17:08.29 & 00:04:20.93 &  & 26.74 & 25.61 & 24.43 & (3) \\
        SSA22-LBG-15 & 22:17:13.07 & 00:05:46.04 &  & 99.00 & 25.10 & 99.00 & (3) \\
        SSA22-LBG-16 & 22:17:12.56 & 00:08:52.28 &  & 99.00 & 25.47 & 99.00 & (3) \\
        SSA22-LBG-17 & 22:17:02.60 & 00:08:26.93 &  & 99.00 & 25.10 & 24.17 & (3) \\
        SSA22-LBG-18 & 22:17:10.27 & 00:04:40.23 &  & 26.85 & 25.05 & 24.13 & (3) \\
        SSA22-LBG-19 & 22:17:04.88 & 00:07:51.98 &  & 27.13 & 25.33 & 99.00 & (3) \\
        Y12LAE-1$^{e}$ & 22:17:01.58 & 00:08:36.71 &  & -- & -- & -- & (4) \\
        Y12LAE-2$^{e}$ & 22:17:03.59 & 00:07:09.87 &  & -- & -- & -- & (4) \\
        \hline
        \multicolumn{8}{c}{Mask-2}\\
        \hline
        LCG-2$^{d}$ & 22:17:23.55 & 00:03:57.61 & 3.311 & 26.64 & 23.44 & 22.54 & (1), (2) \\ 
        SSA22-LBG-09 & 22:17:15.33 & 00:04:44.39 &  & 26.96 & 24.88 & 99.00 & (3) \\
        SSA22-LBG-12 & 22:17:13.60 & 00:06:40.05 &  & 27.02 & 24.42 & 23.56 & (3) \\
        SSA22-LBG-20 & 22:17:34.30 & 00:03:48.72 &  & 24.21 & 23.96 & 24.07 & (3) \\
        SSA22-LBG-21 & 22:17:21.76 & 00:05:58.43 &  & 26.30 & 25.17 & 24.11 & (3) \\
        SSA22-LBG-22 & 22:17:16.07 & 00:04:32.75 &  & 27.42 & 25.50 & 24.34 & (3) \\
        SSA22-LBG-23 & 22:17:29.97 & 00:05:10.33 &  & 27.05 & 25.22 & 24.26 & (3) \\
        SSA22-LBG-24 & 22:17:27.76 & 00:05:27.47 &  & 99.00 & 25.72 & 23.77 & (3) \\
        SSA22-LBG-25 & 22:17:31.10 & 00:06:18.31 &  & 26.85 & 26.03 & 99.00 & (3) \\
        SSA22-LBG-26 & 22:17:26.56 & 00:04:13.81 &  & 27.08 & 24.50 & 24.45 & (3) \\
        SSA22-LBG-27 & 22:17:25.13 & 00:06:14.46 &  & 99.00 & 25.77 & 99.00 & (3) \\
        SSA22-LBG-28 & 22:17:17.72 & 00:05:45.27 &  & 26.67 & 25.33 & 99.00 & (3) \\
        SSA22-LBG-29 & 22:17:32.95 & 00:04:11.91 &  & 99.00 & 25.14 & 23.16 & (3) \\
        SSA22-LBG-30 & 22:17:28.34 & 00:04:45.57 &  & 99.00 & 24.97 & 24.59 & (3) \\
        SSA22-LBG-31 & 22:17:14.44 & 00:05:00.02 &  & 26.68 & 24.66 & 23.30 & (3) \\
        SSA22-LBG-32 & 22:17:22.68 & 00:06:10.35 & 3.906 & 26.47 & 25.19 & 99.00 & (3) \\
        SSA22-LBG-33 & 22:17:12.47 & 00:06:37.72 &  & 99.00 & 25.13 & 99.00 & (3) \\
        SSA22-LBG-34 & 22:17:25.80 & 00:05:48.64 &  & 99.00 & 25.08 & 23.36 & (3) \\
        SSA22-LBG-35 & 22:17:18.00 & 00:04:09.03 &  & 99.00 & 25.41 & 99.00 & (3) \\
        SSA22-LBG-36 & 22:17:20.01 & 00:05:42.18 &  & 26.03 & 25.32 & 23.95 & (3) \\
        SSA22-LBG-37 & 22:17:31.91 & 00:04:16.57 &  & 26.09 & 25.06 & 24.10 & (3) \\
        Y12LAE-3 & 22:17:31.75 & 00:06:17.66 &  & 99.00 & 26.00 & 99.00 & (4) \\
        \hline
        \multicolumn{8}{l}{$^{a}$ When the spectroscopic redshift has been confirmed by the previous works, we show the values.}\\
        \multicolumn{8}{l}{$^{b}$ The entry 99.00 indicates a non-detection at the $1\sigma$ level.}\\
        \multicolumn{8}{l}{$^{c}$ (1) -- \citet{Iwa09ApJ...692.1287I}; (2) -- \citet{Mic17MNRAS.465..316M}; (3) -- Mawatari et al. in preparation; (4) -- \citet{Yam12AJ....143...79Y, Yam12ApJ...751...29Y}.}\\
        \multicolumn{8}{l}{$^{d}$ The \textit{NB359} and \textit{R}-band magnitudes are same as the total magnitude shown in \citet{Mic17MNRAS.465..316M}.}\\
        \multicolumn{8}{l}{$^{e}$ Not listed in SSA22HIT master catalog due to the non-detections in the Subaru/\textit{i$^{\prime}$}-band.}
    \end{tabular}
\end{table*}

\subsubsection{LCG-1 and LCG-2} \label{S2s1s1:lcg}

LCG-1 is reported by \citet{Mic17MNRAS.465..316M} as ``LBG03'' in their paper.
In the LyC image (Subaru/\textit{NB359} filter), two small LyC clumps are observed, separated by $\Delta r \sim 0.8^{\prime\prime}$ \citep[see figure 2 in][]{Mic17MNRAS.465..316M} from the peak of the rest-frame UV continuum (Subaru/\textit{R} filter). We set the MOSFIRE slitlet on the peak of the UV continuum and use the wider slit width due to the complex morphology of LyC and UV continuum. The spectroscopic redshift is $z_{\mathrm{spec, Ly\alpha}} = 3.287$ which is measured from the optical low-resolution spectroscopy for the Ly$\alpha$ emission line \citep{Mic17MNRAS.465..316M}. Due to the blue $NB359 - R$ color, we consider LCG-1 to be the primary high-\fesc\ candidate in our sample.

 LCG-2 is also reported by \citet{Mic17MNRAS.465..316M} as ``LBG04'' in their paper. The LyC image reveals a single small LyC clump, with a spatial offset of $\Delta r \sim 0.8^{\prime\prime}$ from the peak of the rest-frame UV continuum \citep[see figure 2 in][]{Mic17MNRAS.465..316M}. In this case, we set the slitlet to simultaneously cover the peak of the UV continuum and the LyC clump. The spectroscopic redshift is measured by \citet{Ste03ApJ...592..728S}. In \citet{Ste03ApJ...592..728S}, LCG-2 is referred to as ``SSA22b-oD8'' and shows $z_{\mathrm{spec, Ly\alpha}} = 3.323$ and $z_{\mathrm{spec, abs}} = 3.311$. According to \citet{Ste03ApJ...592..728S}, $z_{\mathrm{spec, Ly\alpha}}$ and $z_{\mathrm{spec, abs}}$ are measured from the Ly$\alpha$ emission line and the average of some absorption lines, respectively. Due to the red $NB359 - R$ color, we consider LCG-2 to be the low- \fesc\ candidate of our sample.

\subsubsection{SSA22-LBGs and Y12LAEs} \label{S2s1s2:lbglae}

 The SSA22-LBGs, selected from the SSA22HIT master catalog, all have $i^{\prime} < 26.4$ ($\mathrm{S/N} > 5$) at $z_{\mathrm{phot}} = 2.7$--$3.7$. We set the slitlets on the peak of the UV continuum (Subaru/\textit{i}$^{\prime}$ filter) for each object. The SSA22-LBGs satisfy either the color-color selection (i.e., classical Lyman break technique) or the photometric redshift estimated by using Hyperz \citep{Bol00A&A...363..476B}, or both. For our MOSFIRE observation, we first choose the SSA22-LBGs with blue UV spectral slopes $\beta$ ($\beta < -2.0$) as the high-priority targets from the catalog. After that, we fix the two mask fields so as to maximize the number of high-priority targets. The remaining slitlets are set to other SSA22-LBGs. There is still a margin of slitlets which is used for the three Y12LAEs. Under the SSA22HIT project, some of the SSA22-LBGs are observed with the Deep Imaging Multi-Object Spectrograph \citep[DEIMOS:][]{Fab03SPIE.4841.1657F} mounted on the Keck-II telescope. In cases where the spectroscopic redshift has been successfully measured by the SSA22HIT project, we show the $z_{\mathrm{spec}}$ value in Table \ref{tab2}.

 There are some SSA22-LBGs significantly detected in the LyC image in our sample: 11 objects display a LyC detection at the $3 \sigma$ level ($NB359 < 26.7$). 6 out of the 11 objects also exhibit a blue UV slope $\beta$ ($\beta < - 2.0$), whereas the remaining 5 objects exhibit a red UV slope $\beta$ ($\beta > - 2.0$). Initially, we considered them as possible LCG candidates. However, from our MOSFIRE observations, we cannot identify any emission lines in any of these targets. Therefore, we conclude that these LBGs are not $z \sim 3$ galaxies and that the \textit{NB359} flux is not LyC. Hence, we do not use these object in the following analysis.

\subsection{Rough estimation of LyC escape fraction} \label{S2s3:lyc}

\begin{figure}
    \includegraphics[width=\columnwidth]{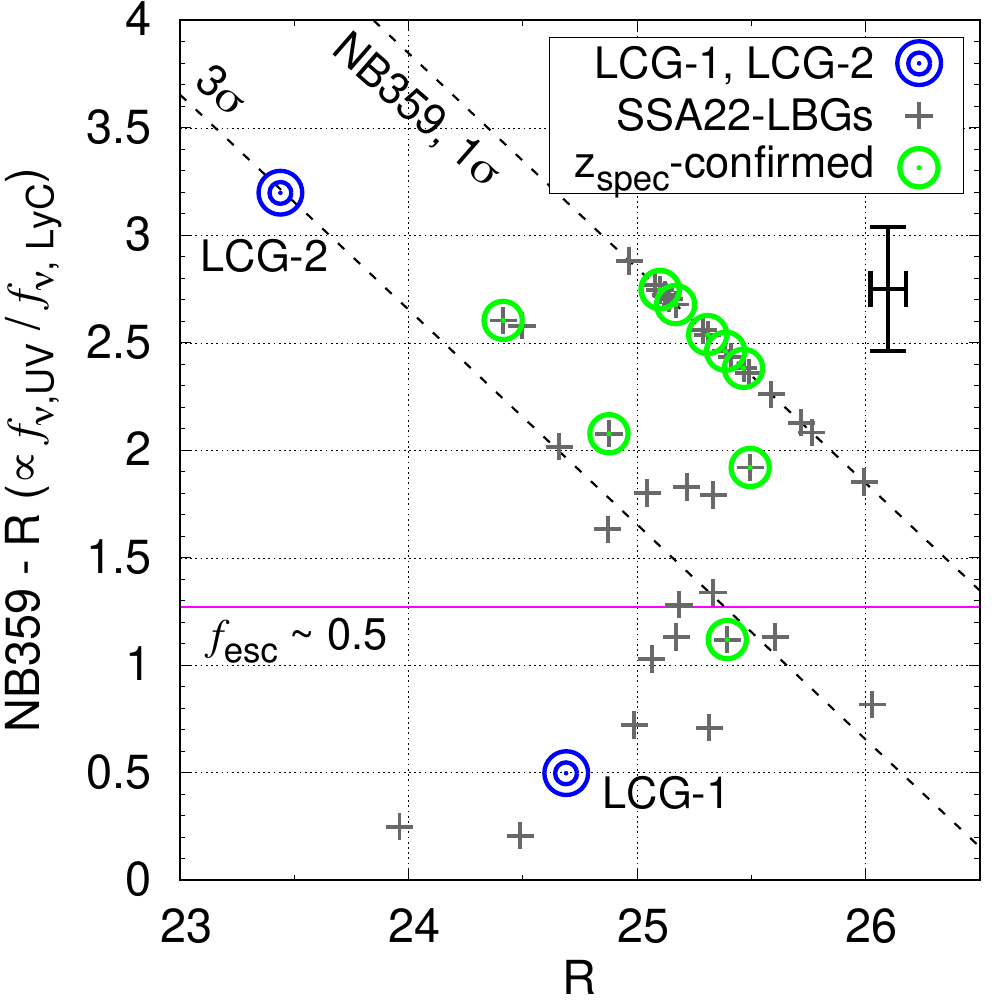}
    \caption{Observed color--magnitude diagram of $NB359 - R$ vs $R$. The blue double circles show our main targets, LCG-1 and LCG-2. The grey crosses denote SSA22-LBGs. The green circles over-plotted on the grey crosses mark the redshift-confirmed SSA22-LBGs, after MOSFIRE spectroscopy. The black error-bars at top right indicate the average error of $NB359 - R$ and $R$. The horizontal magenta line indicates $f_{\mathrm{esc}} \sim 0.5$ under the assumption of a set of standard parameters (see section \ref{S2s3:lyc}).}
    \label{fig1}
\end{figure}

 Fig. \ref{fig1} shows an observed color-magnitude diagram which indicates a rough \fesc\ value and UV flux for our LCGs/SSA22-LBGs. Since the Subaru/\textit{NB359} directly traces LyC photons from galaxies at $z \gtrsim 3.06$, the vertical axis represents the observed flux density ratio $f_{\nu, \mathrm{UV}}/f_{\nu, \mathrm{LyC}}$. The horizontal magenta line indicates $f_{\mathrm{esc}} = 0.5$ assuming a set of standard parameters: IGM transmission $T_{\mathrm{IGM}} = 0.4$ \citep{Ino14MNRAS.442.1805I}, UV dust attenuation $A_{\mathrm{UV}} = 1.67$ \citep{Mic17MNRAS.465..316M}, and the intrinsic LyC-to-UV luminosity ratio $L_{\mathrm{UV}}/L_{\mathrm{LyC}} = 3$, where $L$ is in units of $\mathrm{erg}\, \mathrm{s}^{-1}\, \mathrm{Hz}^{-1}$ \citep[e.g.][]{Ste01ApJ...546..665S, Ino05A&A...435..471I}. In general, the $T_{\mathrm{IGM}}$, $A_{\mathrm{UV}}$, and $L_{\mathrm{UV}}/L_{\mathrm{LyC}}$ values change from galaxy to galaxy. For example, the $L_{\mathrm{UV}}/L_{\mathrm{LyC}}$ value ranges from 1.5 to 5.5 depending on the stellar population age ($0\,\mathrm{Myr}$--$1\,\mathrm{Gyr}$) and the metallicity ($0.05\,Z_{\odot}$--$Z_{\odot}$) in the case of constant star formation history (SFH) and \citet{Sal55ApJ...121..161S} initial mass function (IMF) with $[0.1\,\mathrm{M_{\odot}}$--$100\,\mathrm{M_{\odot}}]$ \citep{Ino05A&A...435..471I}.
As a result, the observed color of $NB359 - R$ for $f_{\mathrm{esc}} = 0.5$ (magenta line in figure \ref{fig1}) changes from $0.5$ to $1.9$; the plausible \fesc\ value of LCG-1 changes to $> 0.5$.
The assumed parameter, $L_{\mathrm{UV}}/L_{\mathrm{LyC}} = 3$, means the relatively younger age (a few $\times 10\,\mathrm{Myr}$) and the lower metallicity ($Z < Z_{\odot}$), which is preferable to LCG-1 due to its blue $NB359 - R$ color.
These parameters ($T_{\mathrm{IGM}}$, $A_{\mathrm{UV}}$, and $L_{\mathrm{UV}}/L_{\mathrm{LyC}}$) should be carefully estimated by the spectral energy distribution (SED) fitting analysis for each object \citep[e.g.][]{Fle19ApJ...878...87F}. In this paper, the rough \fesc\ estimate provided by this figure will be used in an attempt to test the EW(H$\beta$)--$\beta$ method. However, we intend to carry out a more careful \fesc\ estimation for each object in future works by using a $z_{\mathrm{spec}}$ catalog for the SSA22HIT project.

 As indicated by Fig. \ref{fig1}, LCG-1 appears to have $f_{\mathrm{esc}} \gtrsim 0.5$, whereas LCG-2 and other SSA22-LBGs for which $z_{\mathrm{spec}}$ is confirmed by our MOSFIRE observation (green circles over-plotted on grey crosses) appear to have $f_{\mathrm{esc}} < 0.5$. The one exception is SSA22-LBG-08 ($NB359 - R \sim 1.1$).
While this objects seems to have $f_{\mathrm{esc}} \sim 0.5$, the Subaru/\textit{NB359} filter is contaminated by another object close to the SSA22-LBG-08. Therefore, we adopt $f_{\mathrm{esc}}<0.5$ for SSA22-LBG-08 as well.
We note two of the green circles are almost overlapped due to the same \textit{R}-band magnitude and the non-detection of \textit{NB359}, although there are ten $z_{\mathrm{spec}}$-confirmed SSA22-LBGs in Fig. \ref{fig1}.
Some of the other SSA22-LBGs (grey crosses only) also show the significant detection with \textit{NB359} and a high \fesc\ value. However, we are unable to confirm $z_{\mathrm{spec}}$ of the \textit{NB359}-detected SSA22-LBGs due to either a single or no emission line in our MOSFIRE observation. It is therefore likely that the Subaru/\textit{NB359} filter does not trace the LyC emission for these objects and we will not discuss these objects further.

\section{MOSFIRE observations} \label{S3Mo}

\subsection{Observation and data reduction} \label{S3s1Oadr}

 We conduct \textit{K}-band multi-object spectroscopy with the Multi-Object Spectrometer For Infra-Red Exploration \citep[MOSFIRE;][]{McL10SPIE.7735E..1EM, McL12SPIE.8446E..0JM} on the Keck-I telescope. The purpose of these observations is to determine the spectroscopic redshift using strong emission lines such as [\ion{O}{iii}] $\lambda\lambda\, 4959, 5007$, and to measure the H$\beta$ emission line flux. The data was obtained on August 27, 2016 during photometric conditions with seeing 0.4$^{\prime\prime}$--0.6$^{\prime\prime}$. We applied the ABA$^{\prime}$B$^{\prime}$ dithering pattern with a nod amplitude of 3.0$^{\prime\prime}$ (A-B) and 2.4$^{\prime\prime}$ (A$^{\prime}$-B$^{\prime}$). We mainly used slit widths of 0.7$^{\prime\prime}$, which corresponds to a spectral resolution $\mathrm{R} \sim 3600$ in the \textit{K}-band, although a slit width of 0.8$^{\prime\prime}$ was used for LCG-1 to cover the LyC emitting regions (see section \ref{S2s1s1:lcg}).

 We use two slit masks, Mask-1 and Mask-2, for our 42 LCGs/LBGs/LAEs. The field of view covered by the two slit-masks partially overlap, and SSA22-LBG-09 and SSA22-LBG-12 are observed using both slit masks. In addition to our LCGs/SSA22-LBGs targets, we also assign a slitlet on a bright and point-source like object ($K \sim 20$) in each slit mask to assess slit losses. The exposure time of Mask-1 and Mask-2 are 2.45h and 0.5h, respectively. Due to the short exposure time used for Mask-2, we fail to detect emission lines from most of the targets observe in this setting. We only use LCG-2 and SSA22-LBG-22 from Mask-2 for our following analysis.

 The data reduction is performed by using the MOSFIRE data reduction pipeline (DRP)\footnote{\url{https://keck-datareductionpipelines.github.io/MosfireDRP}}, which was developed by the MOSFIRE instrument team \citep{Ste14ApJ...795..165S}. As a result, we obtain reduced 2-D spectra which are flat-fielded, wavelength calibrated, rectified, and sky subtracted. In the wavelength calibration procedure, we search for the best solution for each slitlet from the combination of OH night sky lines and arc lines of a neon lamp. According to \citet{Ste14ApJ...795..165S}, the wavelength of the final 2-D spectra are reduced to the vacuum wavelength and corrected for the heliocentric velocity. Therefore, we do not apply any additional correction to the reduced 2-D spectra. The 1-D spectra with their $1\sigma$ uncertainties are extracted from the reduced 2-D spectra by using the {\small{BMEP}}\footnote{\url{https://github.com/billfreeman44/bmep}} software developed by the MOSFIRE Deep Evolution Field team \citep[MOSDEF;][]{Fre19ApJ...873..102F}.
For the flux calibration, we use a telluric standard A0V star, HIP 80974, and the bright object in each slit mask. The telluric star is observed at similar air mass as the science frames. The total flux of the bright object is calculated from the UKIRT/\textit{K}-band photometry after taking the filter response of the UKIRT/\textit{K}-band into account. In the procedure for the flux calibration, the 1-D spectra are simultaneously corrected for slit losses.

\begin{figure}
    \includegraphics[width=\columnwidth]{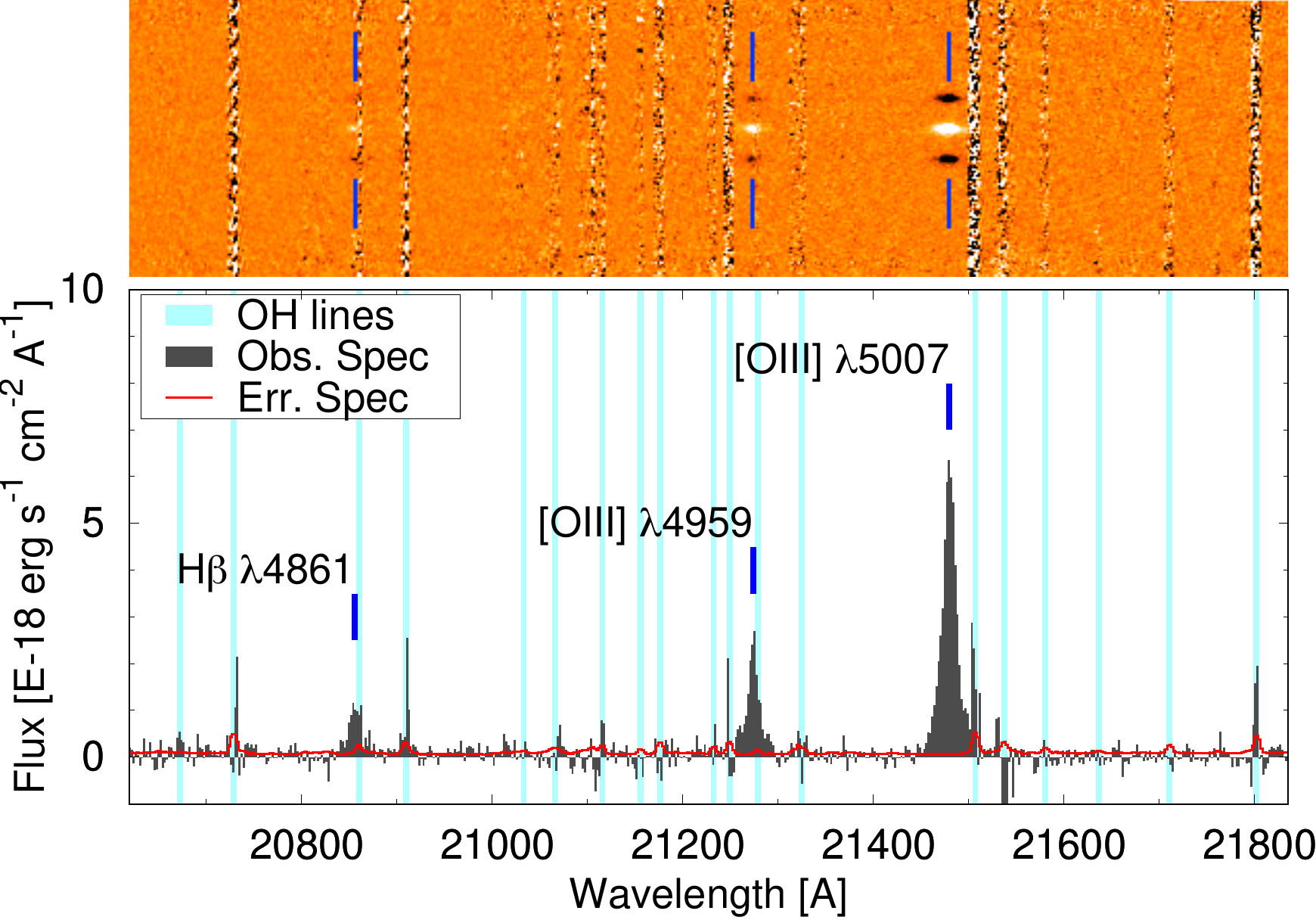}
    \caption{ The reduced 2-D (top) and the extracted 1-D spectra (bottom) of LCG-1. In both panels, the identified emission lines are marked with blue thick lines. In the bottom panel, the filled dark grey histogram and the red lines represent the observed spectrum and the noise spectrum, respectively, in units of $10^{-18}\, \mathrm{erg}\, \mathrm{s}^{-1}\, \mathrm{cm}^{-2}$\, \AA$^{-1}$. The vertical cyan lines indicate the wavelengths of OH night sky lines. }
    \label{fig2}
\end{figure}

 We show the reduced 2-D and the extracted 1-D spectra of LCG-1 in Fig. \ref{fig2} as an example. In this case, we can easily identify some emission lines, and we can extract the 1-D spectrum. In some cases, however, we are unable to identify any emission lines or the continuum from the 2-D spectrum at the object position, and hence refrain from extracting a 1-D spectrum. In Appendix, we summarize all the reduced 2-D spectra and the successfully extracted 1-D spectra.

\subsection{Quick summary of MOSFIRE observations} \label{S3s2qsMo}

\subsubsection{LCG-1 and LCG-2} \label{S3s2s1:lcg}
For LCG-1, we detect the three emission lines ([\ion{O}{iii}] $\lambda \lambda\, 4959, 5007$ and H$\beta$), and then confirm its systemic spectroscopic redshift from [\ion{O}{iii}] $\lambda\, 5007$ to be $z_{\mathrm{spec, sys}} = 3.2890$. This value is consistent with (or slightly larger than) that from the Ly$\alpha$ emission line inferred from optical low-resolution spectroscopy. Although the Ly$\alpha$ emission line may be blueshifted compared with its systemic redshift, we need the optical medium-/high-resolution spectroscopy to conclude the shift. As shown in Fig. \ref{fig2}, part of the H$\beta$ emission line is affected by OH night sky lines. However, we judge their influence on the total line flux to be small because we are able to measure the peak of the H$\beta$ emission line. Moreover, we do not find any other emission lines stemming from lower redshifts. Hence, there is no evidence that the LyC flux inferred for this target from the Subaru/\textit{NB359} image is contaminated by low-$z$ interlopers.

For LCG-2, we detect the three emission lines ([\ion{O}{iii}] $\lambda \lambda\, 4959, 5007$ and H$\beta$). The systemic redshift estimated from [\ion{O}{iii}] $\lambda\, 5007$ is $z_{\mathrm{spec, sys}} = 3.3152$, which is consistent with the redshifts reported by \citet{Ste03ApJ...592..728S} and \citet{Mic17MNRAS.465..316M} since $z_{\mathrm{spec, sys}}$ typically lies somewhere between $z_{\mathrm{spec, Ly\alpha}}$ and $z_{\mathrm{spec, abs}}$ \citep[e.g.][]{Ste10ApJ...717..289S}. Since we do not find any other suspicious emission lines, there is no evidence that the LyC photons detected in the Subaru/\textit{NB359} filter stems from a low-$z$ interloper. The bright absolute UV magnitude of LCG-2 ($M_{\mathrm{UV}} \sim -22.0$) could potentially indicate the existence of the faint AGN in this source. In fact, the observed line width of H$\beta$ is relatively large (${\mathrm{FWHM}}_{\mathrm{raw}} \sim 300\, \mathrm{km}\, \mathrm{s}^{-1}$; Table \ref{ap:tab1}). However, we do not find any other features of AGNs from our MOSFIRE observation. In order to conclude the existence of AGNs, we would need further follow-up observations such as optical spectra covering \ion{C}{iv} $\lambda 1549$ and \ion{He}{ii} $\lambda1640$.

\subsubsection{SSA22-LBGs and Y12LAEs} \label{S3s2s2:lbglae}
For SSA22-LBGs and Y12LAEs, 1--3 emission lines are confirmed for some of the targets after visual inspection of the 2-D images and the 1-D spectrum. There exist six SSA22-LBGs for which we detected the three emission lines ([\ion{O}{iii}] $\lambda \lambda\, 4959, 5007$ and H$\beta$).
For five SSA22-LBGs/Y12LAEs, we confirmed their spectroscopic redshifts through the detection of two emission lines, although their H$\beta$ total flux was not measured (for various reasons -- see next paragraph).
For the rest 29 SSA22-LBGs/Y12LAEs, we detected only a single or no emission line and could not identify their redshifts by our observation itself.
In the following analysis, we use all of six SSA22-LBGs with the detection of three emission lines and only one SSA22-LBG (SSA22-LBG-22) with the detection of two emission lines.

We here describe the details of the five objects with the detection of two emission lines to motivate the reason for using only SSA22-LBG-22 for our analysis. For SSA22-LBG-22, our observation significantly detects [\ion{O}{iii}] $\lambda\, 5007$, and marginally detects [\ion{O}{iii}] $\lambda\, 4959$ ($\mathrm{S/N} \lesssim 5$) due to blending with OH night sky lines. It is in principle possible that the detection of [\ion{O}{iii}] $\lambda\, 4959$ is spurious and SSA22-LBG-22 is actually a lower-z object. However, the spectrum does not contain any further unidentified emission lines and the H$\beta$ wavelength is not affected by the OH night sky lines. Therefore, we consider SSA22-LBG-22 as a robust member of the redshift-confirmed sample.

Three of the five objects are SSA22-LBG-06, 17, and Y12LAE-1. Their H$\beta$ wavelengths are not covered by our observation due to the slitlet configuration. Because we cannot obtain any information about H$\beta$, we do not use these objects. The last of the five objects is SSA22-LBG-08, for which the region covering the expected H$\beta$ wavelength is affected by OH sky lines. Due to the poor constraint on the H$\beta$ emission, we refrain from using SSA22-LBG-08 in our analysis.

\begin{table*}
    \centering
	\caption{ Summary of the observed values used for the EW(H$\beta$)--$\beta$ method. }
	\label{tab3}
    \begin{tabular}{rcccccc}
        \hline
        Name & $z_{\mathrm{spec}}$ $^{a}$ & $m_{K}$ $^{b}$& $\beta$ $^{c}$ & $F_{\mathrm{H}\beta}$ $^{d}$ & $EW_{\mathrm{rest}}$(H$\beta$)$^{e}$ & $f_{\mathrm{esc}}$ $^{f}$\\
        & & & & [$\mathrm{10^{-18}\, erg\> s^{-1}\> cm^{-2}}$] & [\AA] & \\
        \hline
        \multicolumn{7}{c}{Mask-1}\\
        \hline
        LCG-1 & 3.2890 & 23.46 & $-2.44 \pm 0.34$ & 19.82 $\pm$ 0.96 & 104.67 $\pm$ 74.53 & $\leq$ 0.7 \\
        SSA22-LBG-01 & 3.0892 & 24.19 & $-2.35 \pm 0.45$ & 11.72 $\pm$ 0.64 & [38.83 -- 232.07] & 0.0 \\
        SSA22-LBG-02 & 3.3437 & 99.00 & $-2.10 \pm 0.56$ & 10.36 $\pm$ 0.68 & [36.27 -- 170.19] & 0.0 \\
        SSA22-LBG-09 & 2.9768 & 99.00 & $-2.13 \pm 0.32$ & 13.75 $\pm$ 1.64 & [38.68 -- 163.09] & 0.0\\
        SSA22-LBG-10 & 3.2013 & 23.61 & $-1.78 \pm 0.39$ & 14.25 $\pm$ 0.61 & 55.71 $\pm$ 28.80 & 0.0 \\
        SSA22-LBG-12 & 3.1118 & 23.56 & $-2.25 \pm 0.22$ & 27.72 $\pm$ 1.31 & 175.65 $\pm$ 143.71 & 0.0 \\
        SSA22-LBG-16 & 3.1041 & 99.00 & $-1.72 \pm 0.54$ & 4.32 $\pm$ 0.58 & [11.82 -- 52.55] & 0.0 \\
        \hline
        \multicolumn{7}{c}{Mask-2}\\
        \hline
        LCG-2 & 3.3152 & 22.54 & $-1.41 \pm 0.04$ & 53.86 $\pm$ 3.10 & 86.62 $\pm$ 19.62 & $\leq$ 0.5 \\
        SSA22-LBG-22 & 3.3490 & 24.34 & $-1.64 \pm 0.57$ & $<$ 2.19 & $<$ 22.33 & \\
        \hline
        \multicolumn{7}{l}{$^{a}$ $z_{\mathrm{spec}}$ is estimated from the [\ion{O}{iii}] $\lambda\, 5007$ emission line.}\\
        \multicolumn{7}{l}{$^{b}$ The magnitude observed with UKIRT/\textit{K}-band. The values are same as those listed in Table \ref{tab2}.}\\
        \multicolumn{7}{l}{$^{c}$ The UV spectral slope $\beta$ and its uncertainty.}\\
        \multicolumn{7}{l}{$^{d}$ The symbol of $<$ indicates a $3 \sigma$ upper-limit.}\\
        \multicolumn{7}{l}{$^{e}$ [$XX$--$YY$] indicates the permitted range of $EW_{\mathrm{rest}}$(H$\beta$) (see Section \ref{S4s3ewm}).}\\
        \multicolumn{7}{l}{$^{f}$ $f_{\mathrm{esc}}$ is inferred from the EW(H$\beta$)--$\beta$ method.}
    \end{tabular}
\end{table*}

\section{Analysis} \label{S4a}

\subsection{UV spectral slope} \label{S4s1Ussb}

 In this work, we measure the UV spectral slope $\beta$ from the photometry of Subaru/\textit{R}-, \textit{i$^{\prime}$}-, \textit{z$^{\prime}$}-, and HSC/\textit{y}-band filters by using linear weighted least-squares fitting. According to \citet{Fin12ApJ...756..164F} and \citet{Rog13MNRAS.429.2456R}, we apply the following function to the observed photometry,
\begin{equation}
    m(\lambda_{x}) = -2.5 (\beta + 2) \log \lambda_{x} + Const
    \label{eq1}
\end{equation}

\noindent where $\lambda_{x}$ is the effective wavelength of $x$th broad-band filter, $m$($\lambda_{x}$) is the measured magnitude of the $x$th broad-band filter, and ``Const'' is a constant value.
We adopt the best-fit value of the fitting with equation~(\ref{eq1}) on all \textit{R}, \textit{i$^{\prime}$}, \textit{z$^{\prime}$}, and \textit{y}-band magnitudes as the $\beta$ value. The uncertainty on the fitting is also adopted as the uncertainty in the $\beta$ value.
Our sample consists of LBGs at $z \sim 3.2 \pm 0.3$ and LAEs at $z = 3.1$. At these redshifts, the applied broad-band filters of \textit{R}, \textit{i$^{\prime}$}, \textit{z$^{\prime}$}, and \textit{y} in the fitting are optimal for avoiding redshifted strong spectral features such as the Ly$\alpha$ break ($\lambda_{\mathrm{rest}} \sim 1216$\AA) or the Balmer break ($\lambda_{\mathrm{rest}} \sim 3600$\AA). Our $\beta$ values and their uncertainties are listed in Table \ref{tab3}.

\subsection{Emission line measurement} \label{S4s2elfm}

 We adopt a Monte Carlo method for measuring the line profile, the total flux, and the associated uncertainties for each emission line. We first perturb the 1-D spectrum according to its $1 \sigma$ noise spectrum at each pixel (hereafter referred to as a fake spectrum).
We then identify the brightest emission line as [\ion{O}{iii}] $\lambda\, 5007$ from the fake spectrum, and estimate the spectroscopic redshift, $z_{\mathrm{spec}}$, by fitting a single Gaussian profile to the line. If part of the emission line is strongly affected by the OH night sky lines, we mask the pixels in the fitting. The OH night sky lines are identified by using night sky line lists on the MOSFIRE official website\footnote{\url{https://www2.keck.hawaii.edu/inst/mosfire/wavelength_calibration.html}}. We adopt $6.0$\AA\ as the width of each OH line assuming the spectral resolution $\mathrm{R} \sim 3600$. We also calculate the total line flux by integrating the Gaussian profile. We repeat this procedure $10^{4}$ times, and finally obtain the $10^{4}$ measurements of the $z_{\mathrm{spec}}$ value, the FWHM of the line profile, and the total flux from the fake spectra. The mean and standard deviation of the distribution of the measurements are adopted as the best value and its uncertainty, respectively. By using the best $z_{\mathrm{spec}}$ value, we search for the redshifted emission lines of [\ion{O}{iii}] $\lambda\, 4959$ and H$\beta$ from each of the $10^{4}$ fake spectra. When we identify the emission lines, we fit a single Gaussian profile, whose center is fixed based on the best $z_{\mathrm{spec}}$ value, to each line. In a similar way to [\ion{O}{iii}]$ \lambda\, 5007$, we obtain the best FWHM value and the best total flux of [\ion{O}{iii}]$ \lambda\, 4959$ and/or H$\beta$ from the $10^{4}$ fake spectra. For the line fitting, we adopt $\lambda_{\mathrm{rest,vac}} = 5008.240$\AA, $4960.295$\AA, and $4862.683$\AA, which are obtained from the Atomic Line List ver. 2.04\footnote{\url{http://www.pa.uky.edu/~peter/atomic/}}, as a vacuum wavelength of [\ion{O}{iii}] $\lambda\lambda\, 5007, 4959$, and H$\beta$, respectively.
 
 As for SSA22-LBG-22 (the object without the detection of H$\beta$), we measure the $3 \sigma$ upper-limit of the H$\beta$ emission line. On the basis of error propagation, we estimate a $1 \sigma$ uncertainty in the H$\beta$ emission line flux from $\sqrt{\sum \sigma^2}$ where $\sigma$ indicates the $1 \sigma$ uncertainty per spectral element from the noise spectrum. For the range of the summation, we adopt $\lambda = \lambda_{\mathrm{H}\beta,\mathrm{vac}} (1 + z_{\mathrm{spec}}) \pm 3 \sigma_{\mathrm{[\ion{O}{iii}]}}$ where $\sigma_{\mathrm{[\ion{O}{iii}]}}$ is the best standard deviation of [\ion{O}{iii}] $\lambda\, 5007$ calculated by $\sigma_{\mathrm{[\ion{O}{iii}]}} = \mathrm{FWHM}/(2\sqrt{2\ln{2}})$. In the estimation for the upper-limit, we assume that the FWHM of H$\beta$ is similar to that of [\ion{O}{iii}] $\lambda\, 5007$. We consider $\pm 3 \sigma_{\mathrm{[\ion{O}{iii}]}}$ to be optimal since a wider summation range results in an overestimation of the upper limit due to contamination of the OH night sky lines.

 In Table \ref{tab3}, we show the H$\beta$ total flux for LCG-1, LCG-2, and SSA22-LBGs from which we measure the best total flux of H$\beta$. The detail results of the line measurements for our targets are described in Appendix \ref{ap:redd}, and listed in Tables \ref{ap:tab1} and \ref{ap:tab2}.

\subsection{Equivalent width measurement} \label{S4s3ewm}

 The equivalent width (EW) of H$\beta$ is obtained by $EW(\mathrm{H}\beta) = F_{\mathrm{H}\beta}/f_{\lambda, \mathrm{cont}}$ where $F_{\mathrm{H}\beta}$ is the total flux of H$\beta$ and $f_{\lambda, \mathrm{cont}}$ is the continuum flux density at the wavelength of H$\beta$ in units of $\mathrm{erg\> s^{-1}\> cm^{-2}\>}$\AA$^{-1}$.
While we obtain $F_{\mathrm{H}\beta}$ through our MOSFIRE observations, there are no objects detected at a significant level ($> 3 \sigma$) in UKIRT/\textit{K} among the SSA22-LBGs due to the short exposure time of \textit{K} ($K_{3\sigma} \sim 23.5$). However, two SSA22-LBGs are detected at a $2\sigma$ level (SSA22-LBG-10 and SSA22-LBG-12).
For LCG-1, LCG-2, and the two SSA22-LBGs, therefore, we estimate $f_{\lambda, \mathrm{cont}}$ from the \textit{K}-band photometry as an average continuum flux density around the wavelength of H$\beta$ after subtracting the contribution from the [\ion{O}{iii}] and H$\beta$ emission lines to the filter. We simply estimate the uncertainty in EW(H$\beta$) from the uncertainty in the H$\beta$ flux and the \textit{K}-band magnitude on the basis of error propagation. The measured EW(H$\beta$) are summarized in Table \ref{tab3}.

\begin{figure}
    \includegraphics[width=\columnwidth]{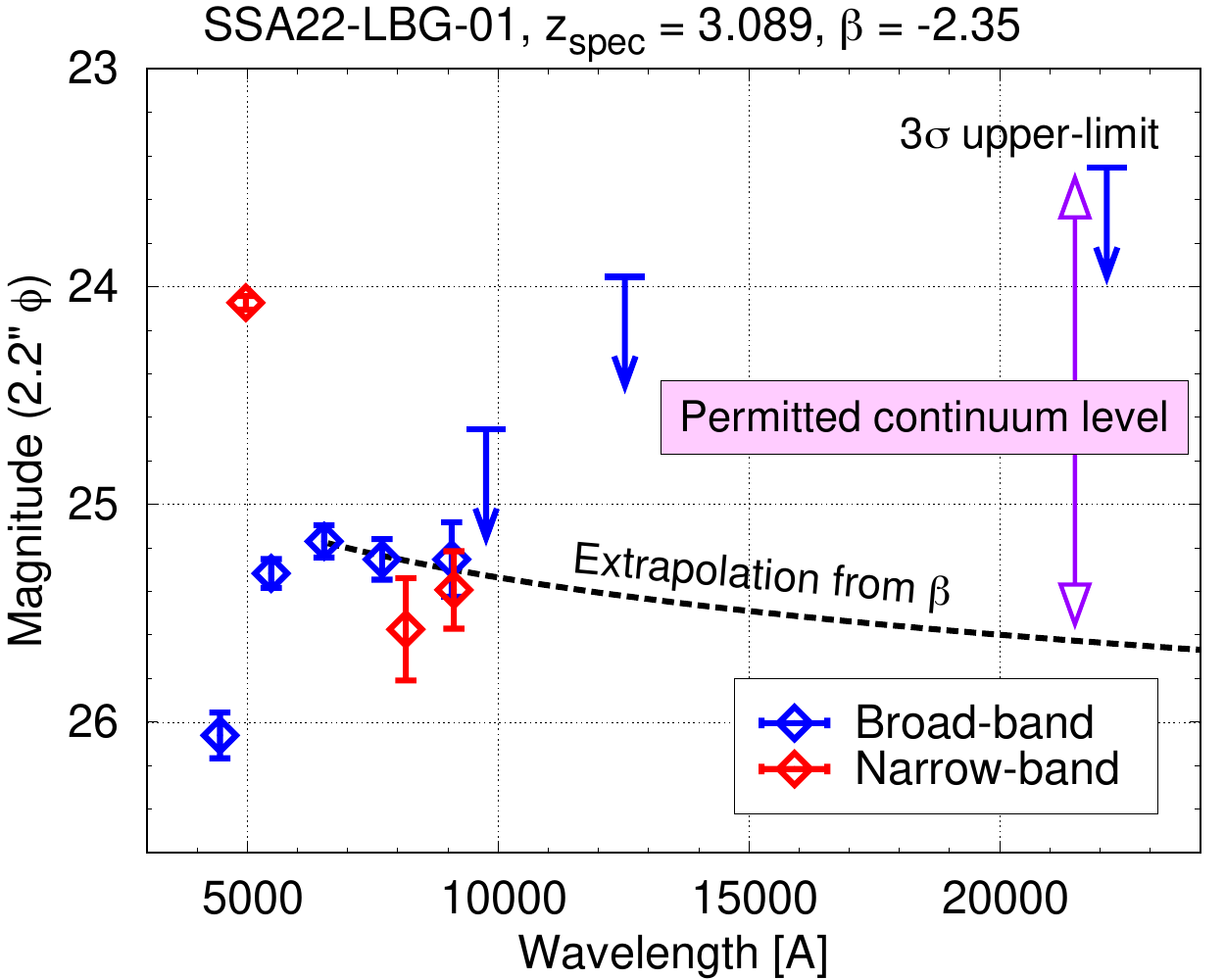}
    \caption{ Illustration for estimating the permitted range of the continuum flux density. As an example, we show the spectral energy distribution of SSA22-LBG-01. The blue diamonds refer to broad-band photometry of \textit{B}, \textit{V}, \textit{R}, \textit{i}$^{\prime}$, \textit{z}$^{\prime}$, \textit{y}, \textit{J}, and \textit{K} from left to right. The red diamonds indicate narrow-band photometry of \textit{NB497}, \textit{NB816}, and \textit{NB912} from left to right. The black dashed line, which shows the best-fit UV spectral slope $\beta$, is extrapolated to longer wavelengths. The purple double arrow indicates the permitted range of the continuum flux density for SSA22-LBG-01. }
    \label{fig3}
\end{figure}

 For other SSA22-LBGs, we regard them as the UKIRT/\textit{K} non-detection sample.
We estimate the permitted range of the continuum flux density as follows. Fig. \ref{fig3} illustrates our idea. The upper limit of the flux density is obtained from the $3 \sigma$ limiting magnitude of the UKIRT/\textit{K}-band image after correcting for the contribution of the [\ion{O}{iii}] and H$\beta$ emission lines to the filter.
The lower limit of the flux density is obtained by extrapolating the UV spectral slope $\beta$ from rest-frame UV to rest-frame optical wavelength. Except for special cases, the continuum flux density in the \textit{K}-band will be similar to or higher than the extrapolation due to the Balmer break which is prominent for old stellar populations. In case of a strong nebular continuum with an extremely young and low-metallicity stellar population, the true continuum flux density can be lower than the extrapolation by $\Delta K \sim 0.3$ at most \citep[e.g.][]{Ino11MNRAS.415.2920I}. Although LyC sources may have such an extreme stellar population, we believe that the extrapolation is a reasonable approximation of the lower limit on the continuum flux density.
By using the upper and lower limits on the continuum flux density, we estimate the permitted range of EW(H$\beta$) for SSA22-LBGs without the \textit{K}-band detection. For SSA22-LBG-22 (the object without the detection of H$\beta$), we estimate the upper limit of EW(H$\beta$) from the $3 \sigma$ upper limit of the H$\beta$ line flux and the lower limit of the continuum flux density.
The permitted range and the upper limit on EW(H$\beta$) are also summarized in Table \ref{tab3}.

\section{Results from the EW(H$\beta$)--$\beta$ diagram} \label{S5r}

\begin{figure}
    \includegraphics[width=\columnwidth]{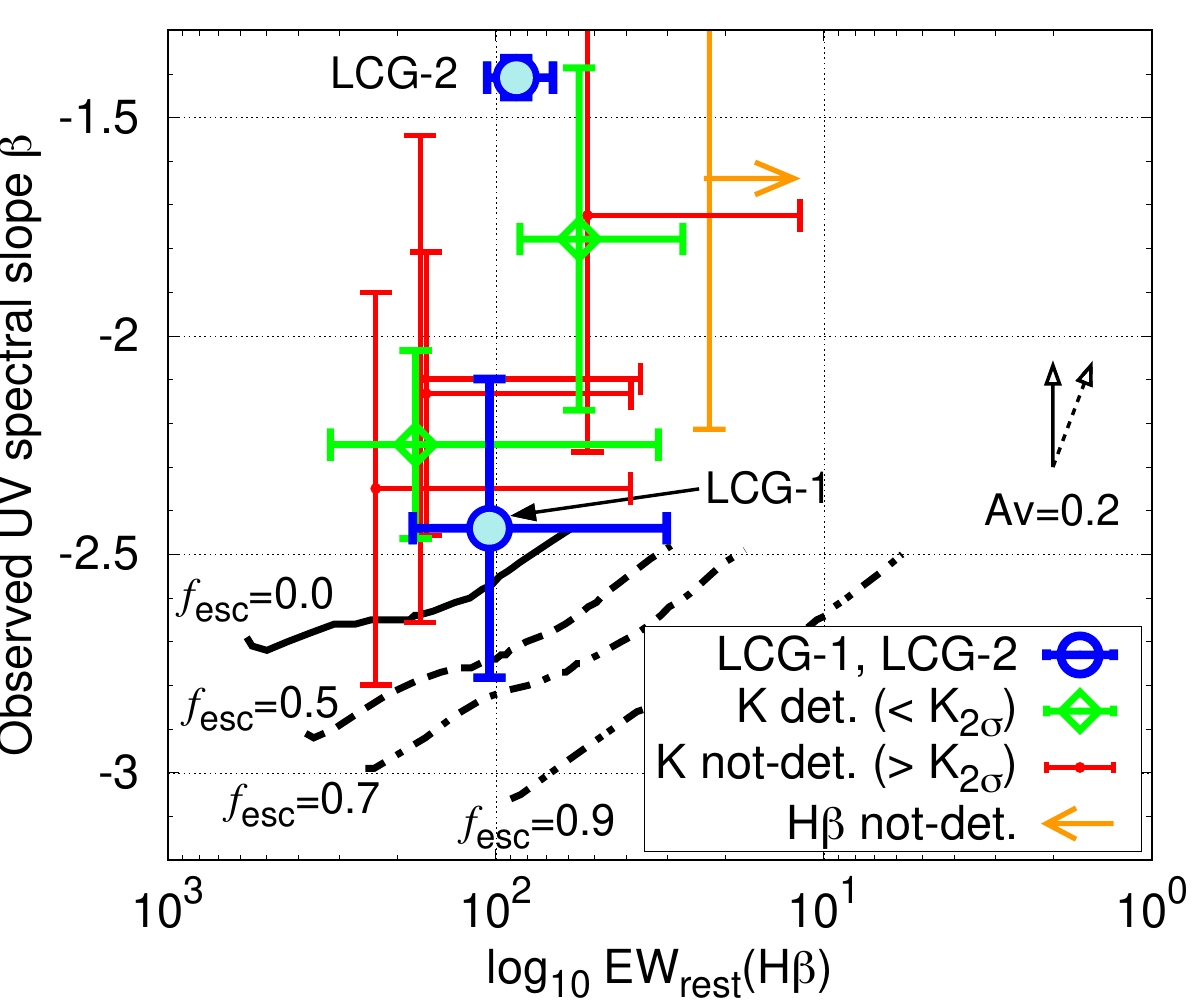}
    \caption{ EW(H$\beta$)--$\beta$ diagram. The x-axis is reversed as the EW(H$\beta$) value increases from right to left. The blue circles filled with cyan show our main targets, LCG-1 and LCG-2. The green diamonds show the SSA22-LBGs with marginal \textit{K}-band detection at $> 2 \sigma$ level (K-LBGs). The red error bars show the SSA22-LBGs which are undetected in the \textit{K}-band at the $2 \sigma$ level (nK-LBGs). The red error bars along with y-axis represent the permitted range of EW(H$\beta$) estimated from the upper/lower limits on continuum flux density (see section \ref{S4s3ewm}). The orange error bar with the arrow shows the SSA22-LBG-22 (the object without the detection of H$\beta$) and indicates the $3\sigma$ upper limit of EW(H$\beta$). The black solid, dashed, dot-dashed, two-dot-dashed lines show the model tracks for radiation-bounded nebulae, $Z=0.004$, and constant star formation history from \citet{Zac13ApJ...777...39Z}. The two black arrows at the right hand represents the influence of the dust attenuation. For LCG-1, the \fesc\ value estimated  by the EW(H$\beta$)--$\beta$ method is roughly consistent with the \fesc\ value estimated from the direct LyC measurement. }
    \label{fig4}
\end{figure}

 Fig. \ref{fig4} shows the result of the EW(H$\beta$)--$\beta$ method to constrain $f_\mathrm{esc}$.
In this diagram, we include the data for LCG-1, LCG-2, and the SSA22-LBGs listed in Table \ref{tab3}.
The blue circles filled with cyan show our main targets, LCG-1 and LCG-2. The green diamonds with error bars represent UKIRT/\textit{K}-detected SSA22-LBGs (hereafter K-LBGs). The red error bars represent SSA22-LBGs, which are undetected in \textit{K} at the $2 \sigma$ level (hereafter nK-LBGs), whereas the horizontal error bars of the nK-LBGs represent the permitted range of EW(H$\beta$) described in Section \ref{S4s3ewm}.
The uncertainty in $\beta$ is indicated by the vertical error bars. For the sake of clarity, the error bars for $\beta$ are placed at the left edge of the permitted range of EW(H$\beta$). Therefore, the red error bars denote the expected regions in which the nK-LBGs are.
The orange arrow with error bars represents SSA22-LBG-22 (the object without the detection of H$\beta$).
The arrow denotes the upper limit on EW(H$\beta$), estimated from the $3 \sigma$ upper limit on the H$\beta$ flux and the lower limit on the continuum flux density.

 In Fig. \ref{fig4}, we also show the model tracks for radiation-bounded nebulae with various \fesc, $Z=0.004$, and constant SFH from \citet{Zac13ApJ...777...39Z}. The model tracks are based on the \textit{Yggdrasil} spectral synthesis code \citep{Zac11ApJ...740...13Z}, which simulates galaxy SEDs (stellar+nebular continuum emission and nebular emission lines) as a function of age for arbitrary SFHs. In \citet{Zac13ApJ...777...39Z} for the model tracks of EW(H$\beta$)-$\beta$, the simple stellar population SED models are generated by Starburst99 \citep{Lei99ApJS..123....3L} with Padova-AGB stellar evolutionary tracks \citep{Vaz05ApJ...621..695V} assuming the \citet{Kro01MNRAS.322..231K} IMF with $[0.1\, \mathrm{M}_{\odot}$--$100\, \mathrm{M}_{\odot}]$. The final galaxy SEDs are estimated by using the stellar SEDs integrated over the SFH as an input in the photoionization code Cloudy \citep{Fer98PASP..110..761F}.
The black solid, dashed, dot-dashed, and two-dot-dashed lines represent the models with $f_{\mathrm{esc}} = 0.0$, $0.5$, $0.7$, and $0.9$, respectively. Each model track changes with the production rate of ionizing photons in a galaxy, namely, the age of the stellar population. The age of the model galaxy increases from bottom left ($1\, \mathrm{Myr}$) to top right ($300\, \mathrm{Myr}$) along with each model track. Nebular continuum is already applied to these model tracks according to the \fesc\ values.
As discussed in detail in \citet{Zac13ApJ...777...39Z, Zac17ApJ...836...78Z}, a different assumption of stellar populations (e.g. a different metallicity, SFH, IMF, and stellar evolutionary model) shifts the model tracks on the EW(H$\beta$)--$\beta$ diagram. While some alternative set of model parameters may be preferable for LCG-1 (and LCG-2/SSA22-LBGs), it is difficult to have a further discussion due to the large uncertainties in EW(H$\beta$) and $\beta$. Therefore, we consider the simple assumption described above.

These models are for cases without any dust attenuation ($A_{V} = 0.0$), even though the observed LCGs/SSA22-LBGs are affected by dust, to various degree.
The impact of dust on the EW(H$\beta$)--$\beta$ diagram depends on the dust attenuation curve and geometry of dust, gas and stars in the target galaxies \citep{Zac13ApJ...777...39Z, Zac17ApJ...836...78Z}.
 The two black arrows on the right-hand side of Fig. \ref{fig4} represent two simple models for dust attenuation effects. In these simple models, we assume the \citet{Cal00ApJ...533..682C} attenuation curve. The black solid arrow is the case where the dust attenuation of the stellar component is the same as the attenuation of the nebular component ($A_{V} = A_{V,\mathrm{neb}}$). Because the dust attenuation of the continuum and the emission line flux are the same, the arrow runs parallel to the y-axis. The black dashed arrow represents an alternative scenario where the dust attenuation of the stellar component is $0.44 \times$ the dust attenuation of the nebular component ($A_{V} = 0.44 A_{V,\mathrm{neb}}$). The relation between $A_{V}$ and $A_{V,\mathrm{neb}}$ has been discussed by e.g. \citet{Erb06ApJ...647..128E, Ksn13ApJ...777L...8K}. According to these studies, the proportionality constant of the relation seems to range between 0.44 and 1.0 for high-$z$ starbursts. Therefore, the two constant values adopted in our simple models represent reasonable assumptions.

 Due to the blue UV spectral slope ($\beta \approx -2.5$) of LCG-1, we can directly compare its observed EW(H$\beta$) and $\beta$ values to the model prediction from \citet{Zac13ApJ...777...39Z} without considering the influence of dust attenuation. When taking the error bars in to account, we find that the EW(H$\beta$)--$\beta$ method indicates $f_{\mathrm{esc}} = 0.0$--$0.7$ for the object LCG-1. As shown in Fig. \ref{fig1}, the direct LyC measurement also predicts a high \fesc\ value, $f_{\mathrm{esc}} \gtrsim 0.5$, for this object. Hence, the two methods give broadly consistent results.
 It is clear, however, that given the observational uncertainties, the EW(H$\beta$)--$\beta$ method would not be able to provide useful constraints on its own, as it is only able to rule out $f_{\mathrm{esc}}>0.7$ for this object.

 For the remaining objects, it is difficult to directly apply the EW(H$\beta$)-$\beta$ method, since their UV slopes are redder and therefore likely more affected by dust. As an example, consider LCG-2, which displays $\beta \approx -1.4$. For this object, the dust-corrected (intrinsic) UV spectral slope, $\beta_{\mathrm{int}}$, would be required for a detailed constraint on \fesc\ using the EW(H$\beta$)-$\beta$ model sequences.
The intrinsic UV slope $\beta_{\mathrm{int}}$ could potentially be derived from  SED fitting, although it would be premature to attempt this at the current time. For LBGs at $z\approx 4$, the SED-fitting analysis of \citet{SY19PASJ...71...51Y} indicates $\beta_{\mathrm{int}} \sim -2.5 \pm 0.3$ for objects with $\beta_{\mathrm{obs}} \sim -1.4$. Since LBGs with the same $\beta$ value at $z=3$ and $z=4$ are, on average, expected to be similar in terms of stellar populations, one could be tempted to shift LCG-2 to $\beta_{\mathrm{int}} \sim -2.5 \pm 0.3$, in which case the 
EW(H$\beta$)--$\beta$ method would suggest $f_{\mathrm{esc}} = 0.0$--$0.5$ . While this is formally consistent with the prediction from the direct LyC measurement ($f_{\mathrm{esc}} \sim 0.0$), this would place the dust-corrected position of LCG-2 in the EW(H$\beta$)--$\beta$ diagram very close to the observed position of LCG-1, which would seem contradictory given that the direct LyC observations indicate significantly different $f_\mathrm{esc}$ for these two objects.
Similar problems plague the EW(H$\beta$) and $\beta_{\mathrm{int}}$ analysis of the other objects in Fig. \ref{fig4} as well; because the other objects also display redder UV slopes (and have larger error bars), it is hard to provide constraints on $f_\mathrm{esc}$ from their positions expected from $\beta_{\mathrm{int}}$ in the EW(H$\beta$)--$\beta$ diagram at this stage. For the $\beta_{\mathrm{int}}$ estimation, it is required to conduct the detailed SED fitting analysis by using deep and multi-band imaging data.

\section{Discussion} \label{S6d}

\subsection{Probing the LyC escape mechanism}

 Throughout the paper, we have implicitly assumed \fesc\ estimated by the EW(H$\beta$)--$\beta$ method to be the same as that estimated from the direct LyC measurement. It is, however, possible that this assumption may break down in the case where the escape of LyC photons is anisotropic.
 
 Because the direct LyC measurement depends on the optical depth of neutral hydrogen along the line-of-sight, the direct method displays a dependence on the viewing angle. The EW(H$\beta$)--$\beta$ method, on the other hand, does not have the same angular dependence. In the scenario of a `radiation-bounded nebula with holes \citep{Zac13ApJ...777...39Z}, also known as the picket-fence model, the LyC photons escape from the galaxy through holes or tunnels of optically thin (or almost ionized) hydrogen gas. However, the distribution of these holes and their sizes may differ across the object, and the ones facing the observer (responsible for the leakage of LyC photons detected in direct observations) may not be representative of the average across the whole galaxy. If this is the case, direct and indirect methods may not necessarily result in comparable \fesc\ estimates. 
 
Indeed, recent observations indicate that the LyC photons from LCGs with high \fesc\ may have escaped through low-density channels of the type envisioned in the radiation-bounded nebula with holes scenario \citep{Van16ApJ...825...41V, Van18MNRAS.476L..15V, Van19MNRAS.tmp.2218V, Izo18MNRAS.478.4851I}. According to \citet{Ver15A&A...578A...7V} and \citet{Beh14A&A...563A..77B}, a triple-peaked Ly$\alpha$ emission profile with a peak at its systemic velocity (or a double-peaked Ly$\alpha$ emission profile with a small peak separation) is predicted for high \fesc\ LCGs in this scenario, and such a spectral feature has indeed been observed in some previous works \citep{Van18MNRAS.476L..15V, Van19MNRAS.tmp.2218V}. Moreover, \citet{Riv19Sci...366..738R} report on the LyC properties and \fesc\ properties of the multiple images from the gravitationally lensed SFG at $z = 2.4$. The individual images of the LyC source in this object are spatially unresolved within the high-resolution \textit{HST} image, which may indicate narrow escape channels.

 If LyC leakage through a radiation-bounded nebular with holes is a commonly occurring phenomenon, the EW(H$\beta$)--$\beta$ method could become an important tool for probing the ``angle-averaged'' (hereafter global) escape fraction.
As mentioned in Section \ref{S5r}, the EW(H$\beta$)--$\beta$ method does display a dependence on dust (the dust attenuation curve and the geometry of dust, gas, and stars), and hence on the viewing angle.
However, this angular dependence on dust is relatively weaker than the angular dependence of the direct LyC measurements on neutral hydrogen because the dust opacity in the optical/UV wavelength is much smaller than the opacity of neutral hydrogen for LyC photons.
On the other hand, the dust opacity for LyC photons in ionized gas can be important (\citealt{Chi18A&A...616A..30C, Ste18ApJ...869..123S, Gaz20A&A...639A..85G}; see also \citealt{ Ino01ApJ...555..613I,Ino01AJ....122.1788I,Ino02ApJ...570..688I})\footnote{The cross-section of neutral hydrogen to LyC is $\sigma_{\mathrm{HI}} = 6.3 \times 10^{-18}\, \mathrm{cm^{2}}$. The cross-section of dust is $\sigma_{\mathrm{dust}} = \pi \times 10^{-10}\, \mathrm{cm^{2}}$ assuming the radius of dust grain $r = 0.1\, \mu$m and the dust absorption coefficient $Q = 1$. The Dust-to-Hydrogen (atomic $+$ molecular) mass ratio is $D = (m_{\mathrm{dust}} N_{\mathrm{dust}}) / (\mu m_{\mathrm{H}} N_{\mathrm{H}}) = (m_{\mathrm{dust}} N_{\mathrm{dust}}) / (\mu m_{\mathrm{H}} N_{\mathrm{HI}} / X_{\mathrm{HI}})$, where $\mu$ is the mean molecular weight and $X_{\mathrm{HI}}$ is the neutral fraction of hydrogen. Assuming the dust mass density $\rho = 3\, \mathrm{g}/\mathrm{cm}^{3}$ and $\mu = 1.4$, the Dust-to-Hydrogen mass ratio is $D = 5.37 \times 10^{9} X_{\mathrm{HI}} (N_{\mathrm{dust}} /N_{\mathrm{HI}})$. Consequently, the ratio of the opacities of HI and dust is $k_{\mathrm{HI}}/k_{\mathrm{dust}} = (\sigma_{\mathrm{HI}}  N_{\mathrm{HI}}) / (\sigma_{\mathrm{dust}} N_{\mathrm{dust}}) \sim 100\, X_{\mathrm{HI}}/D$. According to \citet{Dra07ApJ...663..866D}, the Dust-to-Hydrogen mass ratio of Milky Way is $7.3 \times 10^{-3}$, which indicates $k_{\mathrm{HI}}/k_{\mathrm{dust}} \sim 1.5 \times 10^{4}\,X_{\mathrm{HI}}$. Although the dust opacity for LyC is negligible in neutral gas, the opacity is not negligible in ionized regions ($X_{\mathrm{HI}} << 1$).}.
In any case, the angular dependence on dust is much weaker when considering SFGs with little dust attenuation like LCG-1 which displays the blue UV slope $\beta$. When assessing the role of galaxies in the reionization of the Universe, it is in fact the global \fesc\ that matters, not \fesc\ in the direction of the observer (which would be measured by the direct LyC measurements). 
 
 By combining direct measurements of LyC escape (which trace optically thin holes or regions) with indirect ones (which may potentially provide a better estimate of the global $f_{\mathrm{esc}})$ for the same objects, it may be possible to quantitatively assess the anisotropy of the leakage, at least in the case of SFGs at $z = 3$--$4$ and for lower-redshift analogs.
The ultimate goal is to understand the LyC escape mechanism of SFGs in the epoch of reionization ($z > 6$) that are predicted to have a little amount of dust attenuation. Therefore, it is worth investigating the difference of the direct and indirect measurements of LyC escape for the lower-redshift analogs of the reionization-epoch SFGs.
It should be stressed, however, that one can only hope to detect very extreme cases of LyC leakage ($f_{\mathrm{esc}} \gtrsim 0.5$) using the  EW(H$\beta$)--$\beta$ method. Similar indirect techniques that make use of a wider set of spectroscopic data could in principle do better \citep{Jen16ApJ...827....5J, Giri20}.

\subsection{Uncertainties in EW(H$\beta$)--$\beta$ method}

 Here, we discuss the observational effects that dominate the uncertainties in the EW(H$\beta$)--$\beta$ method, and how the errors may potentially be reduced in the future. In our analysis, the errors on EW(H$\beta$) and $\beta$ are dominated by the uncertainties in the broad-band photometry at the longer wavelengths, namely, in the Subaru/\textit{z$^{\prime}$}, \textit{y}, and UKIRT/\textit{K}-band filters. Since the spectroscopic H$\beta$ flux is well determined (with a typical uncertainty of $\lesssim 10$\%; Table \ref{tab3}), the uncertainty on EW(H$\beta$) is attributed to the large uncertainty in the continuum flux density estimated from the UKIRT/\textit{K}-band photometry. 
 
 For blue UV spectral slopes ($\beta < -2.0$), the  $z^{\prime}$- and $y$-band fluxes become fainter than those of the $R$- and $i^{\prime}$-bands. Therefore, the uncertainty on the \textit{z$^{\prime}$}- and \textit{y}-band photometry is critical for the $\beta$ estimation. If we were to obtain much deeper images of $z^{\prime}$-, $y$-, and $K$-band in the SSA22 field, the EW(H$\beta$)--$\beta$ would be easier to apply. 
 
A similar issue may also be important for attempts to apply the EW(H$\beta$)--$\beta$ method to SFGs at $z > 6$. Since strong nebular emission lines are expected for SFGs at $z>6$ \citep[e.g.][]{Har18ApJ...859...84H}, emission line identification may be easily accomplished by future instruments such as \textit{JWST}. However, if the H$\beta$ continuum is undetected in the spectroscopic data, as would be expected for very faint sources and/or observations at high spectral resolution, deep imaging observations may nonetheless be important to measure the continuum flux.

 The complicated effects of dust on the EW(H$\beta$)--$\beta$ diagram also adds substantial uncertainty in the application of the  EW(H$\beta$)--$\beta$ method for estimating $f_\mathrm{esc}$. The observational data in Fig. \ref{fig4} are affected by dust, whereas the model sequences from \citet{Zac13ApJ...777...39Z} are for \textit{intrinsic} EW(H$\beta$) and $\beta$ prior to dust attenuation. Both the attenuation curve and the amount of dust attenuation are critical parameters in the exercise for applying the right corrections to EW(H$\beta$) and $\beta$. As mentioned in section \ref{S5r}, if the dust attenuation applied for the nebular component is different from that for the stellar component, the observational data points of the EW(H$\beta$)--$\beta$ diagram shift diagonally from  bottom left to top right as the amount of the dust attenuation is increased.
 
For LCG-1 (and some SSA22-LBGs), however, the observed UV spectral slope is $\beta \approx -2.5$, which is quite blue compared to the average $\beta$ value of LBGs/LAEs at similar redshifts \citep[$\beta \sim -1.7$ or $-2.0$ for $M_{\mathrm{UV}} \sim -21.0$ at $z \sim 3$; e.g.][]{Bou09ApJ...705..936B, San19arXiv191002959S}. Due to the strong dependence of $\beta$ on the dust attenuation \citep[e.g.][]{Meu99ApJ...521...64M}, these blue UV spectral slopes $\beta$ indicate that dust reddening must be small. Therefore, the observed EW(H$\beta$) and $\beta$ are expected to be close to their intrinsic values, at least for LCG-1 and possibly also for some of the other SSA22-LBGs in our sample.

\section{Conclusion} \label{S7c}
 In this work, we examine the validity of the EW(H$\beta$)--$\beta$ method proposed by \citet{Zac13ApJ...777...39Z} to indirectly estimates the escape fraction of LyC photons from individual galaxies. For this purpose, we conduct \textit{K}-band multi-object spectroscopy with Keck/MOSFIRE to measure the H$\beta$ emission line flux of LCGs, LBGs, and LAEs at $z \sim 3.0$--$3.5$ in the SSA22 field.
 Thanks to the unique Subaru/\textit{NB359} filter, we can also directly detect the LyC flux from these objects and assess the associated escape fraction  \fesc. Finally, we compare the outcomes of these two methods and discuss the usefulness of combining direct and indirect methods for probing the astrophysical mechanism that allow LyC photons to escape from galaxies. 
 
Our main conclusions can be summarized as:
\begin{enumerate}
\item
 We reconfirm the spectroscopic redshift and measure the H$\beta$ emission line flux from 2 LCGs and 6 SSA22-LBGs. The spectroscopic redshifts of our main targets, LCG-1 and LCG-2, are $z_{\mathrm{spec}} = 3.2890$ and $3.3152$, in good agreement with the Ly$\alpha$ redshift values reported by \citet{Mic17MNRAS.465..316M}.
 
\item
 When plotting our 8 LCGs/SSA22-LBGs onto the EW(H$\beta$)--$\beta$ diagram, we find LCG-1 to be suitably located for assessing $f_\mathrm{esc}$ using the \citet{Zac13ApJ...777...39Z} method. Based on the model predictions, we infer $f_{\mathrm{esc}} = 0.0$--$0.7$ for this object. While this is  broadly consistent with the escape fraction estimated from the directly detected LyC flux of this object ($f_\mathrm{esc}\approx 0.5$), the large error bars on $\beta$ and EW(H$\beta$) prevent us from setting strong quantitative constraints on the agreement. The remaining objects display UV continuum slopes $\beta$ that are too red to allow any meaningful comparison without the application of dust corrections to the $\beta$ slopes.

\item
We discuss the possibility that direct and indirect LyC leakage estimates could return discrepant estimates of $f_\mathrm{esc}$ in the case of anisotropic LyC leakage, as in the case where LyC leakage happens through a radiation-bounded nebula with holes. In this scenario, the direct LyC measurements trace the covering fraction along the line of sight. This quantity may differ from the angle-averaged, ``global'' escape fraction that matters for the role of galaxies in the reionization of the Universe, and which could potentially be better probed through indirect $f_\mathrm{esc}$-estimation techniques like the EW(H$\beta$)-$\beta$ method. Hence, improved comparisons between these two techniques for SFGs at $z = 3$--$4$ and low-$z$ analogs could provide useful constraints on the anisotropy of LyC leakage. 

\end{enumerate}

\section*{Acknowledgements}
 SY and AKI was supported by JSPS KAKENHI Grant Namber 17H01114.
 TH was supported by Leading Initiative for Excellent Young Researchers, MEXT, Japan. EZ acknowledges funding from the Swedish National Space Board. The data presented herein were obtained at the W. M. Keck Observatory, which is operated as a scientific partnership among the California Institute of Technology, the University of California and the National Aeronautics and Space Administration. The Observatory was made possible by the generous financial support of the W. M. Keck Foundation.
 This work is based on data collected at Subaru Telescope, which is operated by the National Astronomical Observatory of Japan.
 The UKIDSS project is defined in \citet{Law07MNRAS.379.1599L}. UKIDSS uses the UKIRT Wide Field Camera \citep[WFCAM;][]{Cas07A&A...467..777C}. The photometric system is described in \citet{Hew06MNRAS.367..454H}, and the calibration is described in \citet{Hod09MNRAS.394..675H}. The pipeline processing and science archive are described in Irwin et al (2009, in prep) and \citet{Ham08MNRAS.384..637H}. We use UKIDSS data release 10.
 Data analysis was in part carried out on the Multi-wavelength Data Analysis System operated by the Astronomy Data Center (ADC), National Astronomical Observatory of Japan.
 The authors wish to recognize and acknowledge the very significant cultural role and reverence that the summit of Maunakea has always had within the indigenous Hawaiian community. We are most fortunate to have the opportunity to conduct observations from this mountain.
 
\section*{Data availability}
The data underlying this article will be shared on reasonable request to the corresponding author.



\bibliographystyle{mnras}
\bibliography{BibTex_SY20} 




\appendix

\section{Reduced 2-D images and 1-D spectra of all our targets} \label{ap:redd}
 We here show all the reduced 2-D spectra image and the successfully extracted 1-D spectra in Fig. \ref{ap:fig1cont1}. In cases where the 1-D spectrum is not extracted, we only show the 2-D spectral image. In Tables \ref{ap:tab1} and \ref{ap:tab2}, we show the results of the flux measurements. The following is brief comments on some SSA22-LBGs listed in Table \ref{ap:tab2} (objects with the detection of just a single emission line), i.e. SSA22-LBG-05, SSA22-LBG-13, and Y12LAE-3.
 
 \textit{SSA22-LBG-05}. This object is observed by Keck/DEIMOS under the SSA22HIT project. The spectroscopic redshift is $z_{\mathrm{spec, Ly}\alpha} = 3.112$ which is measured from the DEIMOS observation for the Ly$\alpha$ emission line. If assuming the single emission line detected by our MOSFIRE observation is [\ion{O}{iii}] $\lambda 5007$, the spectroscopic redshift is $z_{\mathrm{spec, [\ion{O}{iii}]}} = 3.1154$.
 Since the spectroscopic redshift is consistent in the observations, SSA22-LBG-05 is actually a redshift-confirmed object. However, the H$\beta$ wavelength is not covered by our observation due to the MOSFIRE slitlet configuration. Therefore, this object is not used for our analysis.

 \textit{SSA22-LBG-13}. This object is observed by Keck/DEIMOS under the SSA22HIT project. The spectroscopic redshift is $z_{\mathrm{spec, Ly}\alpha} = 2.869$. If assuming the single emission line detected by our MOSFIRE observation is [\ion{O}{iii}] $\lambda 5007$ or H$\alpha$ ($\lambda_{\mathrm{vac}} = 6564.61$), the spectroscopic redshift is $z_{\mathrm{spec, [\ion{O}{iii}]}} = 3.7580$ or $z_{\mathrm{spec, H}\alpha} = 2.6299$, respectively. The spectroscopic redshift is not consistent in the observations. Therefore, this object is still a unidentified object.

 \textit{Y12LAE-3}. This object is taken from the LAE candidates at $z_{\mathrm{phot}} = 3.1$ studied in \citet{Yam12AJ....143...79Y}, although the narrow-band excess was below their criteria for the robust LAE sample. If assuming the single emission line detected by our MOSFIRE observation is [\ion{O}{iii}] $\lambda 5007$, the spectroscopic redshift is $z_{\mathrm{spec, [\ion{O}{iii}]}} = 3.1147$. Since the spectroscopic redshift is consistent with the photometric redshift, Y12LAE-3 is a new redshift-confirmed object by our observation. However, the H$\beta$ wavelength is not covered by our observation due to the MOSFIRE slitlet configuration. Therefore, this object is not used for our analysis.

\begin{table*}
    \centering
	\caption{ Summary of flux measurement for LCG-1, LCG-2, and the SSA22-LBGs with the detection of $> 2$ emission lines. }
	\label{ap:tab1}
    \begin{tabular}{rcccccccccc}
        \hline
        Name & $z_{\mathrm{spec}}$ $^{a}$ & \multicolumn{3}{c}{[\ion{O}{iii}] $\lambda 5007$} & \multicolumn{3}{c}{[\ion{O}{iii}] $\lambda 4959$} & \multicolumn{3}{c}{H$\beta$ $^{f}$} \\
        & & $\lambda^{b}$ & FWHM$^{c}$ & Total Flux$^{d}$ & $\lambda^{b,e}$ & FWHM$^{c}$ & Total Flux$^{d}$ & $\lambda^{b,e}$ & FWHM$^{c}$ & Total Flux$^{d}$ \\
        \hline
        \multicolumn{11}{c}{Mask-1}\\
        \hline
        LCG-1 & 3.2890 & 21480.1 & 248.0 & 112.57 $\pm$ 0.77 & 21274.5 & 233.9 & 38.05 $\pm$ 0.87 & 20855.8 & 282.7 & 19.82 $\pm$ 0.96 \\ SSA22-LBG-01 & 3.0892 & 20479.9 & 145.1 & 72.91 $\pm$ 0.68 & 20283.9 & 155.1 & 23.99 $\pm$ 1.11 & 19884.7 & 161.8 & 11.72 $\pm$ 0.64 \\
        SSA22-LBG-02 & 3.3437 & 21754.0 & 205.7 & 60.36 $\pm$ 0.86 & 21545.8 & 245.5 & 21.77 $\pm$ 1.13 & 21121.8 & 211.4 & 10.36 $\pm$ 0.68 \\
        SSA22-LBG-06 & 3.1098 & 20583.0 & 144.7 & 15.78 $\pm$ 1.24 & 20386.0 & 182.6 & 5.24 $\pm$ 0.51 & 19984.8 & -- & -- \\
        SSA22-LBG-08 & 3.3327 & 21699.3 & 131.6 & 22.34 $\pm$ 0.64 & 21491.6 & 112.5 &  6.26 $\pm$ 0.45 & 21068.6 & -99 & -99 \\
        SSA22-LBG-09 & 2.9768 & 19916.5 & 179.8 & 121.30 $\pm$ 0.96 & 19725.9 & 184.0 & 40.07 $\pm$ 0.99 & 19337.7 & 276.3 & 13.75 $\pm$ 1.64 \\
        SSA22-LBG-10 & 3.2013 & 21040.9 & 214.2 & 44.14 $\pm$ 1.08 & 20839.4 & 223.6 & 16.68 $\pm$ 0.85 & 20429.3 & 188.8 & 14.25 $\pm$ 0.61 \\
        SSA22-LBG-12 & 3.1118 & 20592.8 & 172.3 & 171.40 $\pm$ 0.88 & 20395.7 & 171.1 & 58.25 $\pm$ 0.68 & 19994.3 & 166.4 & 27.72 $\pm$ 1.31 \\
        SSA22-LBG-16 & 3.1041 & 20554.2 & 105.1 & 13.24 $\pm$ 1.16 & 20357.5 & 167.1 & 4.42 $\pm$ 0.42 & 19956.8 & 156.7 & 4.32 $\pm$ 0.58 \\
        SSA22-LBG-17 & 3.0734 & 20400.3 & 166.3 & 51.37 $\pm$ 0.75 & 20205.0 & 186.9 & 18.78 $\pm$ 1.10 & 19807.4 & -- & -- \\
        Y12LAE-1 & 3.0824 & 20445.6 & 100.9 & 8.14 $\pm$ 0.38 & 20249.8 & 173.7 & 4.56 $\pm$ 0.56 & 19851.3 & -- & -- \\
        \hline
        \multicolumn{11}{c}{Mask-2}\\
        \hline
        LCG-2 & 3.3152 & 21611.3 & 309.5 & 140.57 $\pm$ 3.11 & 21404.4 & 335.4 & 54.60 $\pm$ 4.41 & 20983.2 & 340.9 & 53.86 $\pm$ 3.10 \\
        SSA22-LBG-09 & 2.9772 & 19918.7 & 178.8 & 117.95 $\pm$ 2.40 & 19728.0 & 201.8 & 39.83 $\pm$ 3.05 & 19339.8 & -- & -- \\
        SSA22-LBG-12 & 3.1116 & 20591.9 & 210.4 & 136.47 $\pm$ 2.19 & 20394.8 & 118.1 & 26.19 $\pm$ 1.05 & 19993.5 & -- & -- \\
        SSA22-LBG-22 & 3.3490 & 21780.9 & 122.2 & 16.37 $\pm$ 1.24 & 21572.4 & 113.2 & 6.86 $\pm$ 1.44 & 21147.9 &  & $<$ 2.19 \\
        \hline
        \multicolumn{11}{l}{$^{a}$ $z_{\mathrm{spec}}$ is estimated from the [\ion{O}{iii}] $\lambda 5007$ emission line. }\\
        \multicolumn{11}{l}{$^{b}$ Center of Gaussian function in units of \AA.}\\
        \multicolumn{11}{l}{$^{c}$ FWHM of Gaussian function in units of $\mathrm{km\> s^{-1}}$. FWHM is not corrected for the instrumental broadening.}\\
        \multicolumn{11}{l}{$^{d}$ Total flux in units of $\mathrm{10^{-18}\, erg\> s^{-1}\> cm^{-2}}$.}\\
        \multicolumn{11}{l}{$^{e}$ The center wavelength of [\ion{O}{iii}] $\lambda 4959$ and H$\beta$ are calculated from $z_{\mathrm{spec}}$ and their vacuum wavelength. They are fixed in the flux measurement. }\\
        \multicolumn{11}{l}{$^{f}$ ``--'' indicates that the predicted wavelength is the outside of the wavelength coverage of our MOSFIRE observation.}\\
        \multicolumn{11}{l}{``-99'' indicates that the predicted wavelength is strongly affected by OH night sky lines.} \\
        \multicolumn{11}{l}{The symbol of ``<'' indicates a $3 \sigma$ upper-limit.}
    \end{tabular}
\end{table*}

\begin{table*}
    \centering
	\caption{ Summary of flux measurement for SSA22-LBGs with the detection of a single emission line. }
	\label{ap:tab2}
    \begin{tabular}{rcccl}
        \hline
        Name & \multicolumn{3}{c}{Unidentified emission line} & Comment \\
        & $\lambda^{a}$ & FWHM$^{b}$ & Total Flux$^{c}$ &  \\
        \hline
        \multicolumn{5}{c}{Mask-1}\\
        \hline
        SSA22-LBG-04 & 22757.1 & 195.7 & 45.37 $\pm$ 1.69 & Line profile looks like [\ion{O}{ii}] $\lambda \lambda 3726,3729$ doublet. \\
        SSA22-LBG-05 & 20611.2 & 182.5 & 32.32 $\pm$ 0.54 & Observed by Keck/DEIMOS under SSA22HIT project. \\
        SSA22-LBG-07 & 21148.2 & 235.9 & 59.75 $\pm$ 0.79 & \\
        SSA22-LBG-13 & 23828.9 & 73.1 & 12.19 $\pm$ 2.57 & Observed by Keck/DEIMOS under SSA22HIT project. \\
        SSA22-LBG-15 & 19414.4 & 281.1 & 15.09 $\pm$ 1.73 & \\
        \hline
        \multicolumn{5}{c}{Mask-2}\\
        \hline
        SSA22-LBG-26 & 19370.5 & 201.1 & 52.54 $\pm$ 3.21 & \\
        SSA22-LBG-31 & 21693.0 & 159.2 & 27.70 $\pm$ 1.93 & \\
        Y12LAE-3 & 20607.5 & 174.9 & 21.39 $\pm$ 1.15 & Suppose line is [\ion{O}{iii}] $\lambda 5007$, $z_{\mathrm{spec}}$ is consistent with $z_{\mathrm{phot}}$ of Y12LAEs.\\
        \hline
        \multicolumn{5}{l}{$^{a}$ Center of Gaussian function in units of \AA.}\\
        \multicolumn{5}{l}{$^{b}$ FWHM of Gaussian function in units of $\mathrm{km\> s^{-1}}$. FWHM is not corrected for the instrumental broadening.}\\
        \multicolumn{5}{l}{$^{c}$ Total flux in units of $\mathrm{10^{-18}\, erg\> s^{-1}\> cm^{-2}}$.}\\
    \end{tabular}
\end{table*}

\begin{figure*}
    \includegraphics[width=2\columnwidth, bb=30 668 552 814]{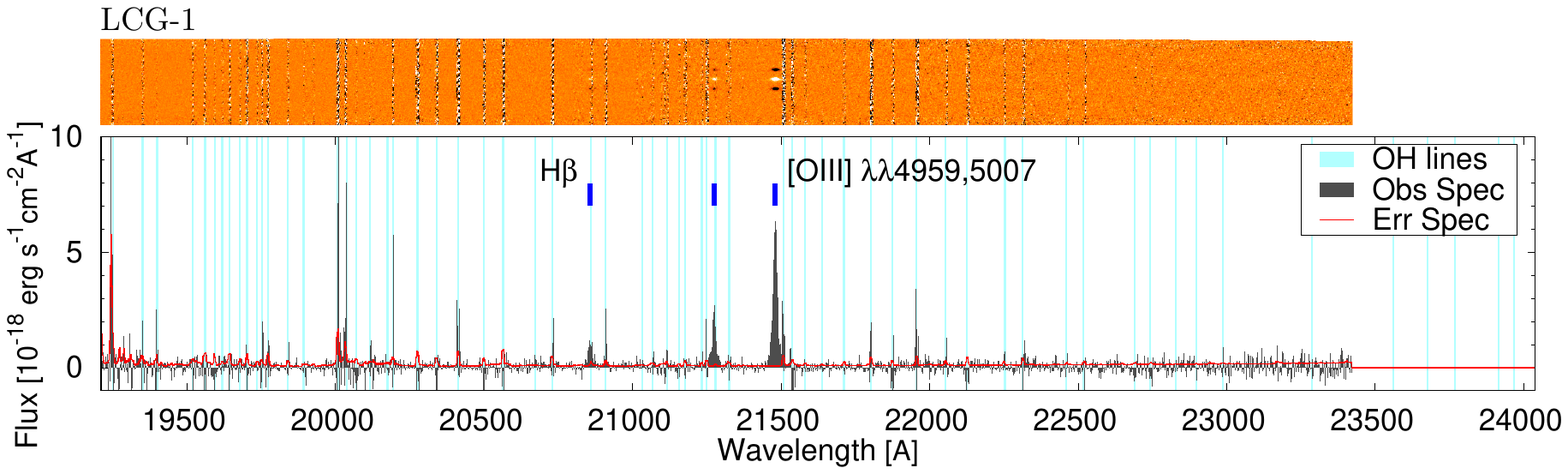}\\
    
    \vspace{10pt}
    \includegraphics[width=2\columnwidth, bb=30 668 552 814]{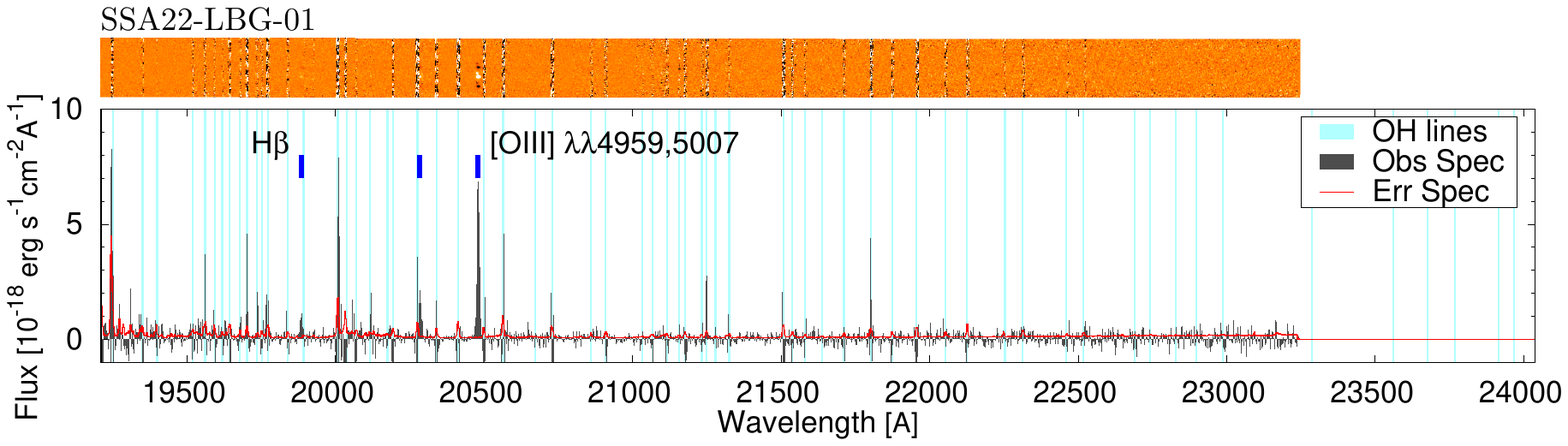}\\
    
    \vspace{10pt}
    \includegraphics[width=2\columnwidth, bb=30 668 552 814]{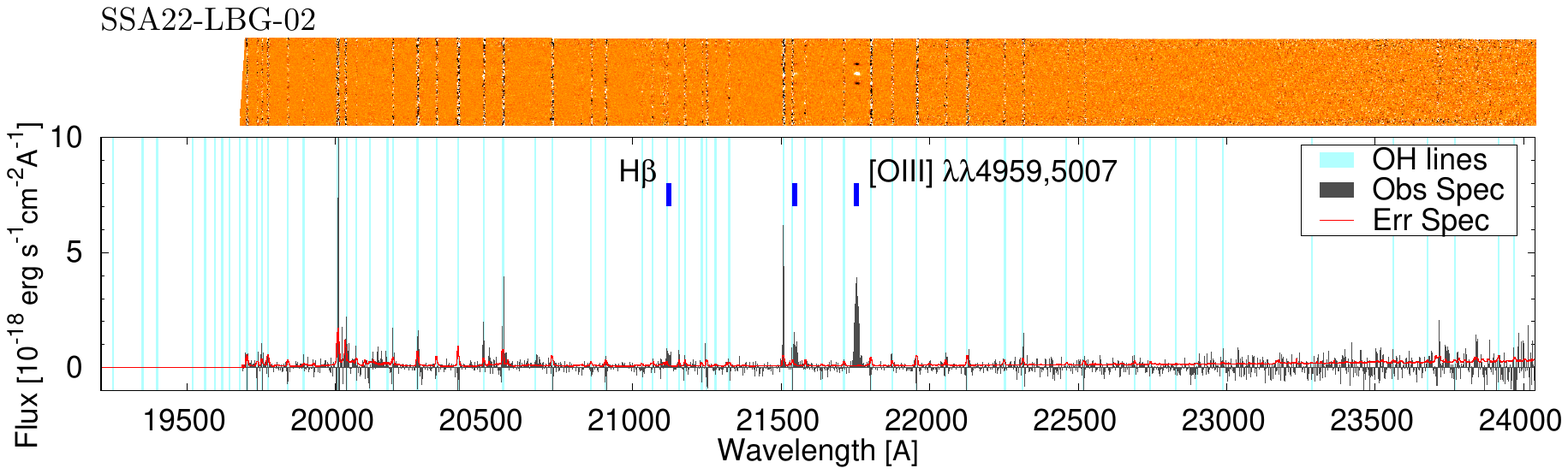}\\
    
    \vspace{10pt}
    \includegraphics[width=2\columnwidth, bb=30 784 552 814]{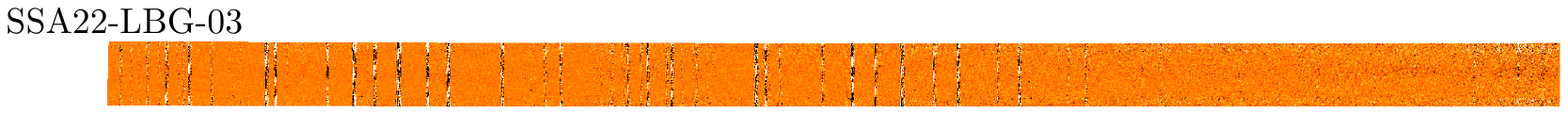}\\
    
    \vspace{10pt}
    \includegraphics[width=2\columnwidth, bb=30 668 552 814]{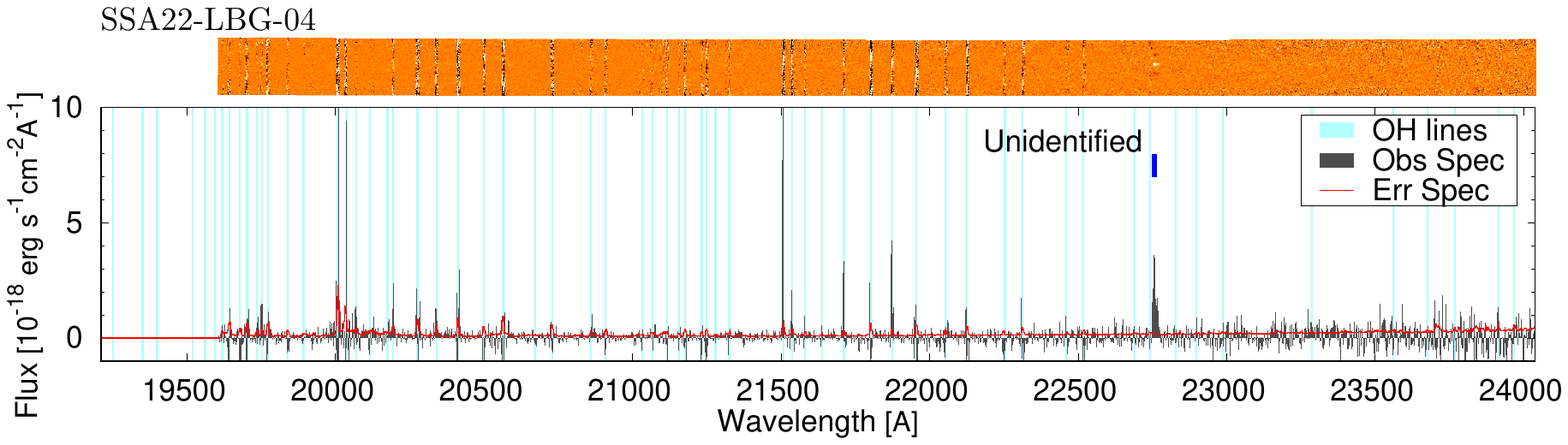}\\
    
    \caption{ Reduced 2-D images and 1-D spectra. The detected lines are marked by blue thick lines.}
    \label{ap:fig1cont1}
\end{figure*}

\begin{figure*}
    \includegraphics[width=2\columnwidth, bb=30 668 552 814]{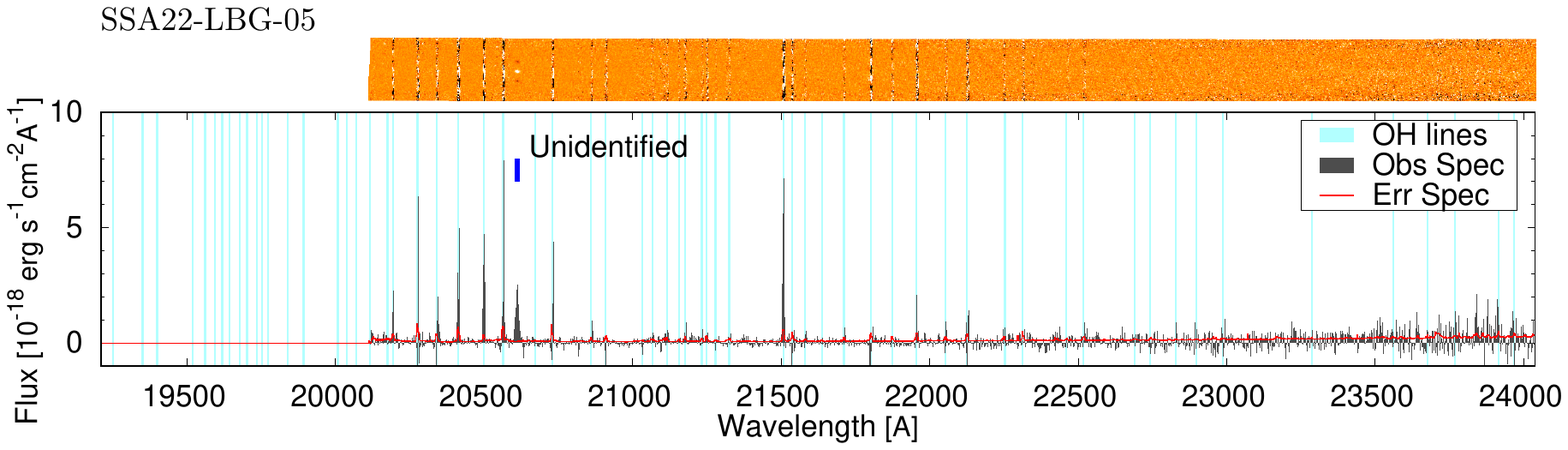}\\
    
    \vspace{10pt}
    \includegraphics[width=2\columnwidth, bb=30 668 552 814]{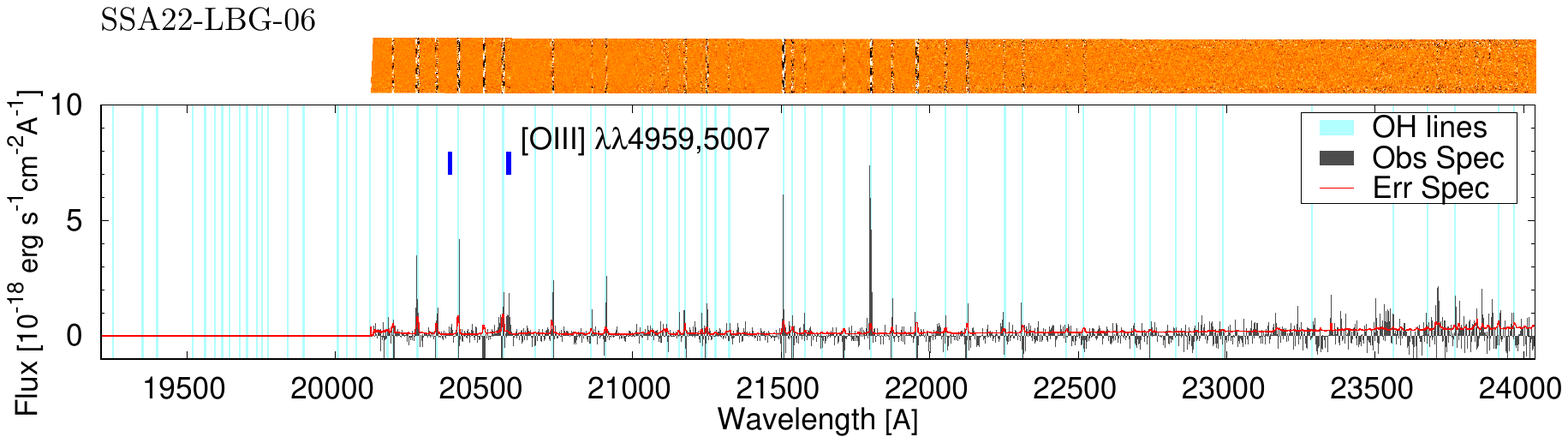}\\
    
    \vspace{10pt}
    \includegraphics[width=2\columnwidth, bb=30 668 552 814]{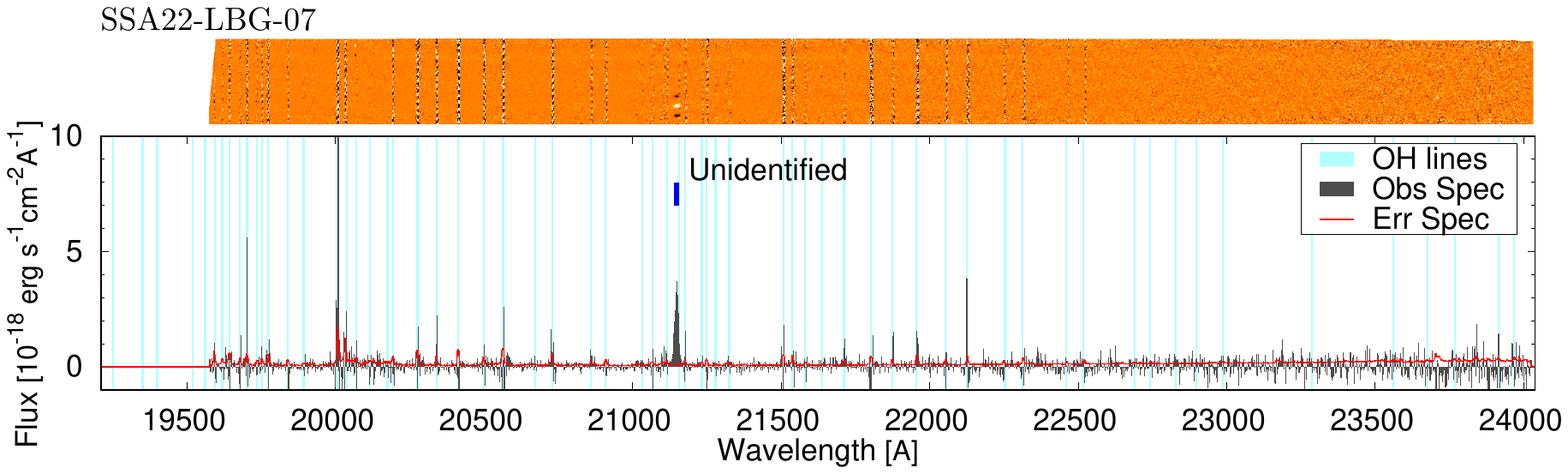}\\
    
    \vspace{10pt}
    \includegraphics[width=2\columnwidth, bb=30 668 552 814]{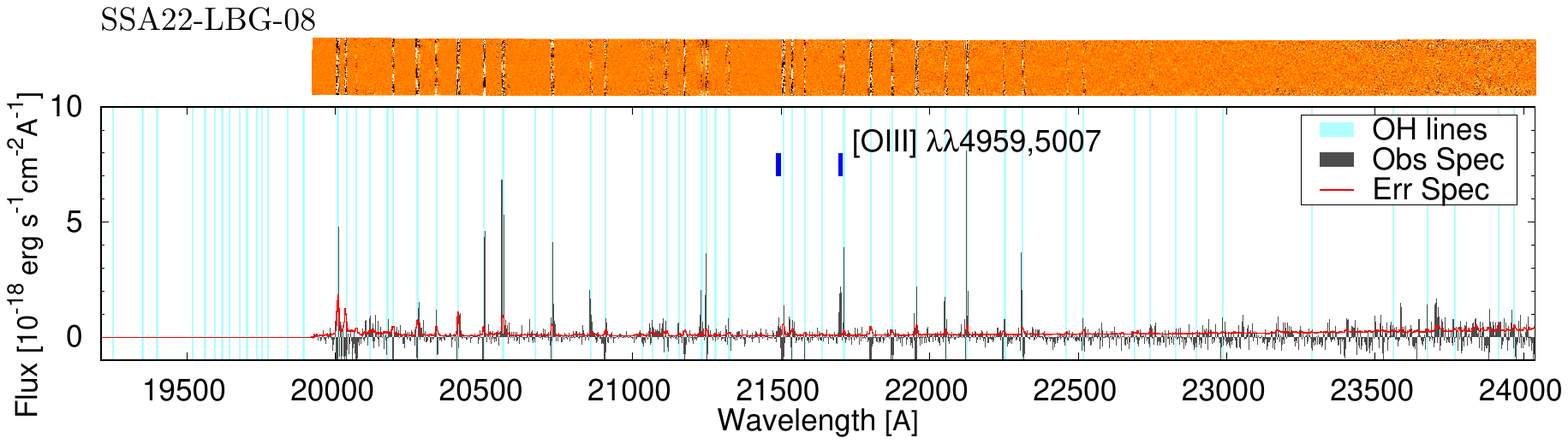}\\
    
    \contcaption{ Reduced 2-D images and 1-D spectra. }
    \label{ap:fig1cont2}
\end{figure*}

\begin{figure*}
    \includegraphics[width=2\columnwidth, bb=30 668 552 814]{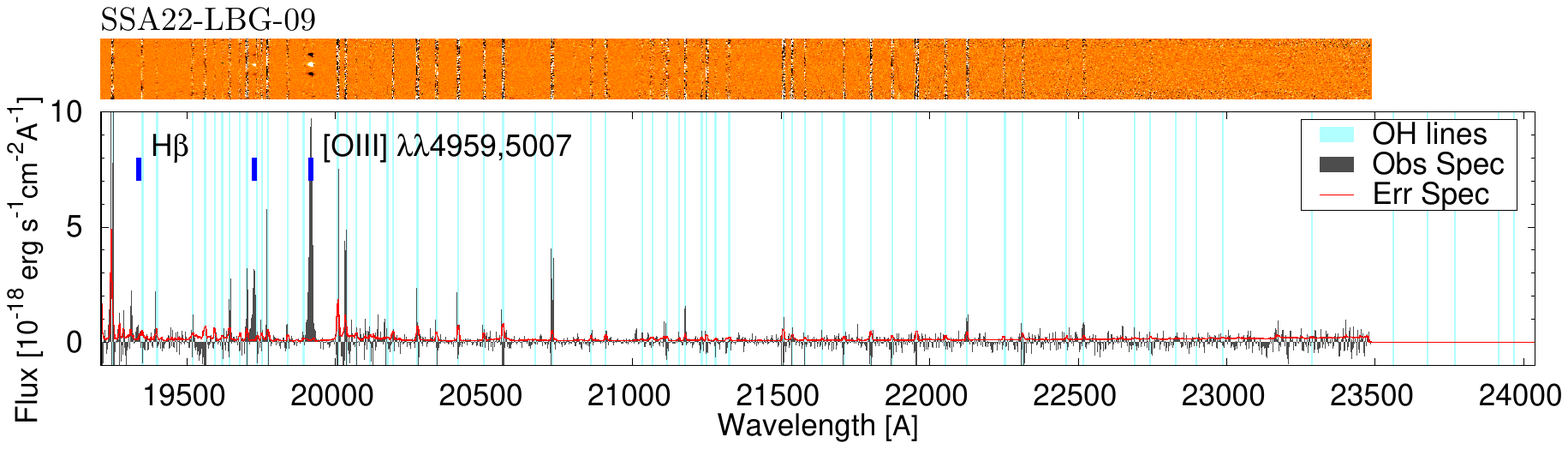}\\
    
    \vspace{10pt}
    \includegraphics[width=2\columnwidth, bb=30 668 552 814]{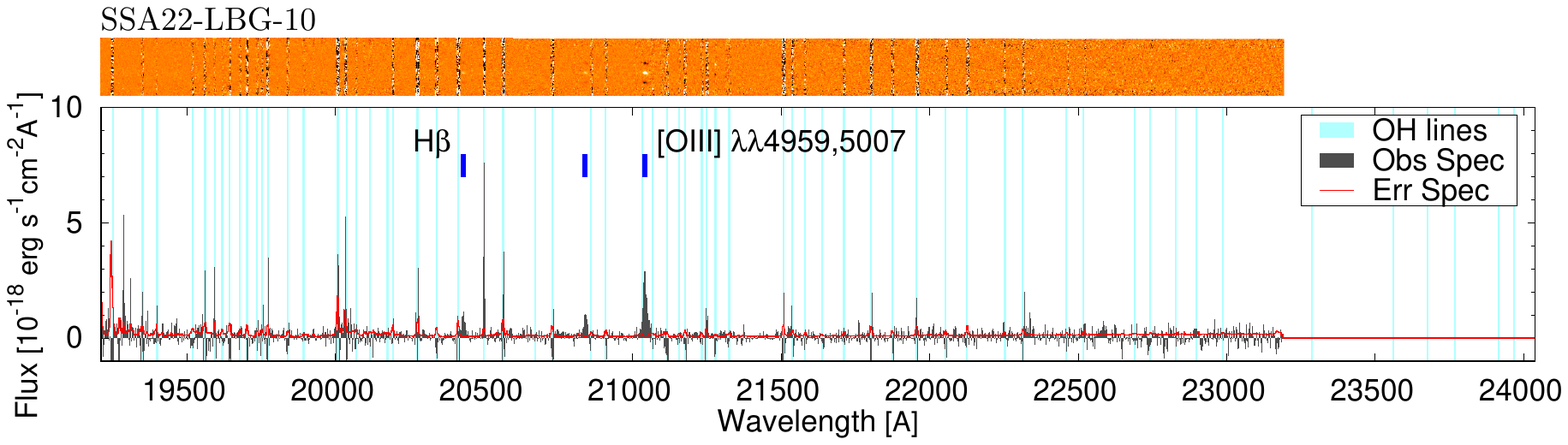}\\
    
    \vspace{10pt}
    \includegraphics[width=2\columnwidth, bb=30 789 552 814]{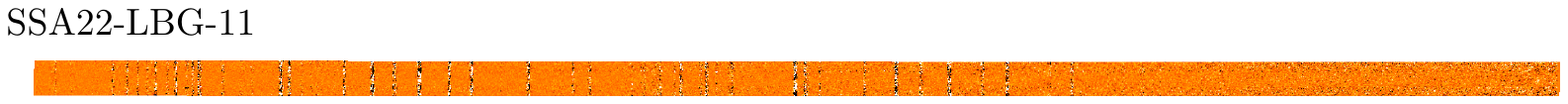}\\
    
    \vspace{10pt}
    \includegraphics[width=2\columnwidth, bb=30 668 552 814]{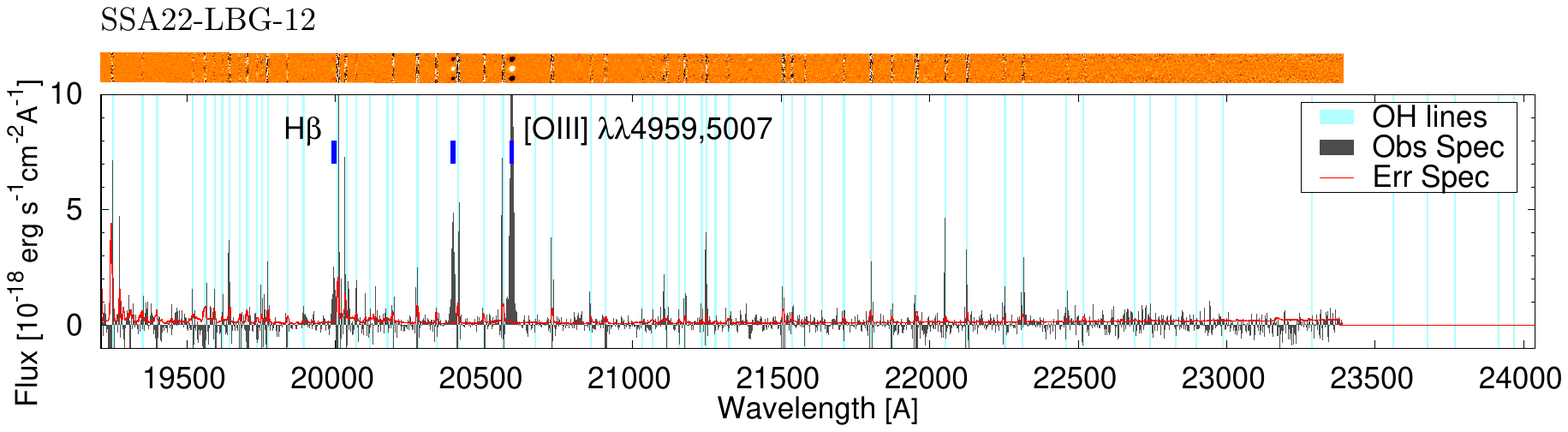}\\
    
    \vspace{10pt}
    \includegraphics[width=2\columnwidth, bb=30 668 552 814]{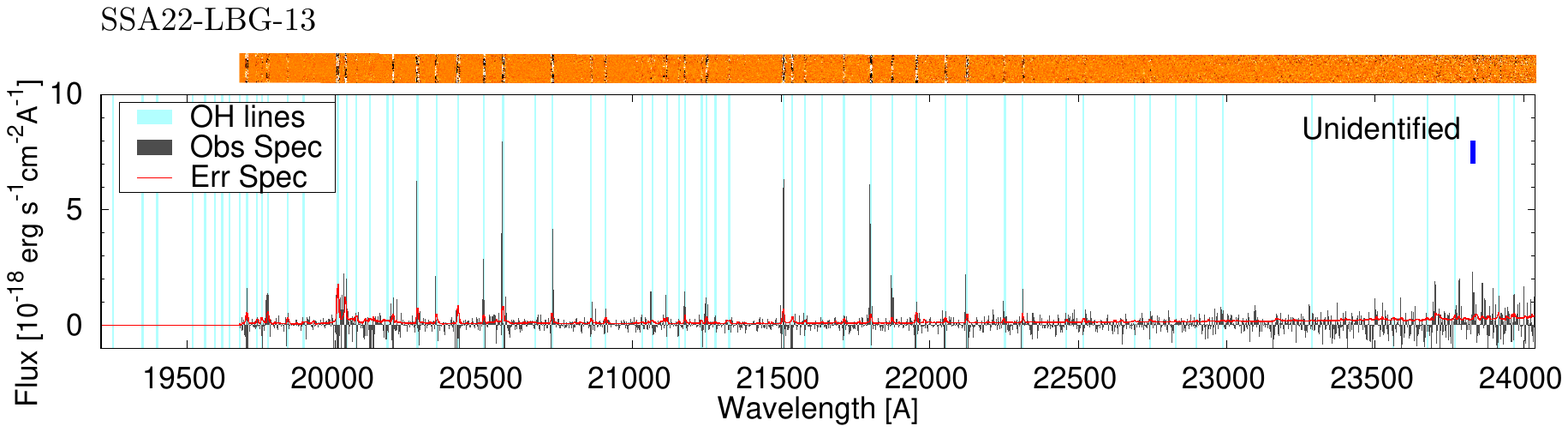}\\
    
    \contcaption{ Reduced 2-D images and 1-D spectra. }
    \label{ap:fig1cont3}
\end{figure*}

\begin{figure*}
    \includegraphics[width=2\columnwidth, bb=30 789 552 814]{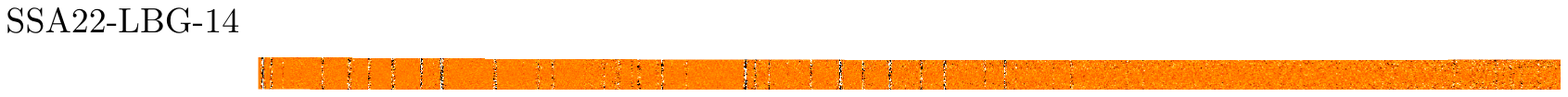}\\
    
    \vspace{10pt}
    \includegraphics[width=2\columnwidth, bb=30 668 552 814]{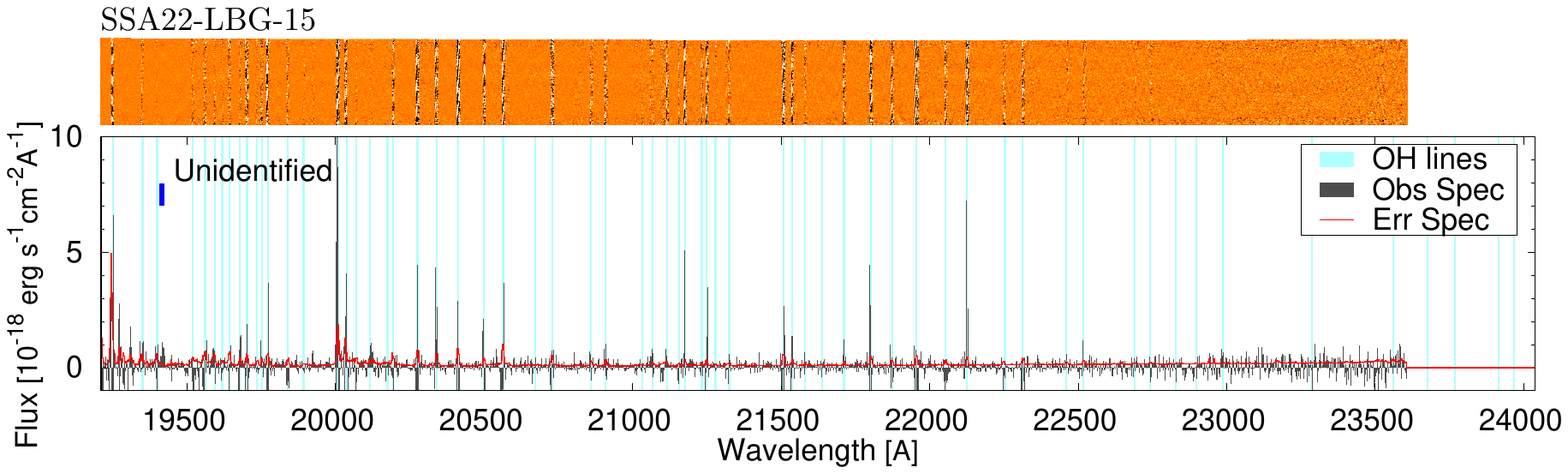}\\
    
    \vspace{10pt}
    \includegraphics[width=2\columnwidth, bb=30 668 552 814]{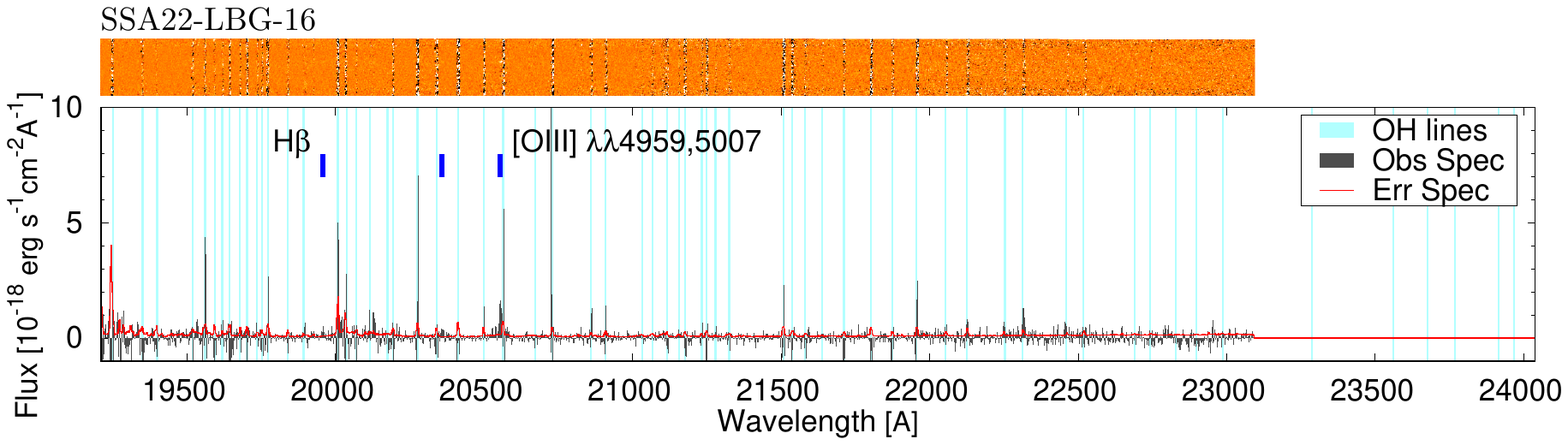}\\
    
    \vspace{10pt}
    \includegraphics[width=2\columnwidth, bb=30 668 552 814]{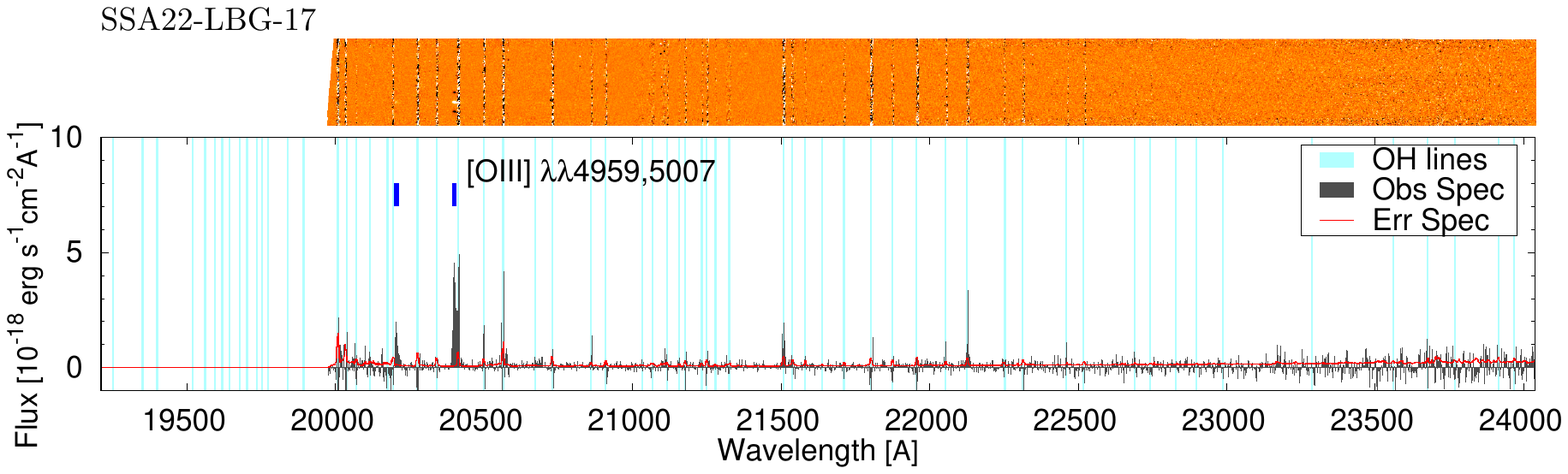}\\
    
    \vspace{10pt}
    \includegraphics[width=2\columnwidth, bb=30 784 552 814]{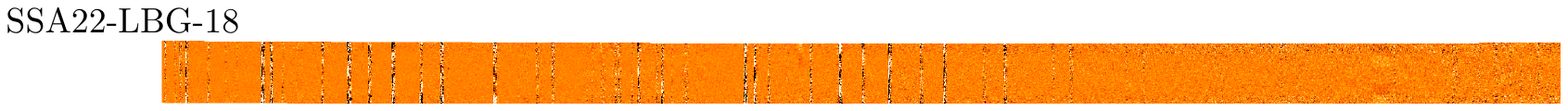}\\
    
    \vspace{10pt}
    \includegraphics[width=2\columnwidth, bb=30 784 552 814]{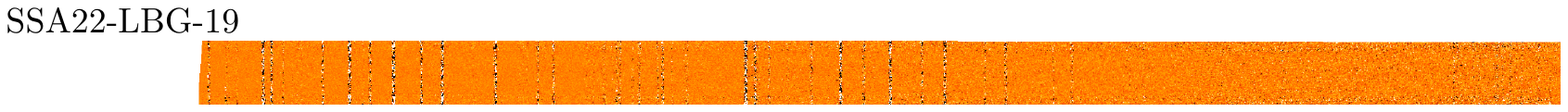}\\
    
    \contcaption{ Reduced 2-D images and 1-D spectra. }
    \label{ap:fig1cont4}
\end{figure*}

\begin{figure*}
    \includegraphics[width=2\columnwidth, bb=30 668 552 814]{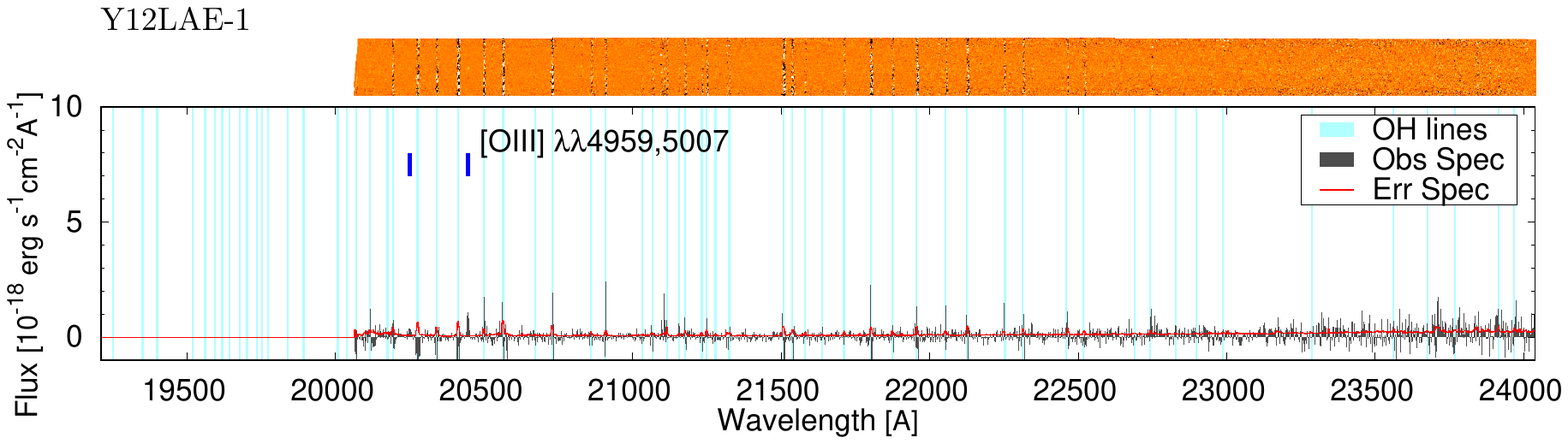}\\
    
    \vspace{10pt}
    \includegraphics[width=2\columnwidth, bb=30 789 552 814]{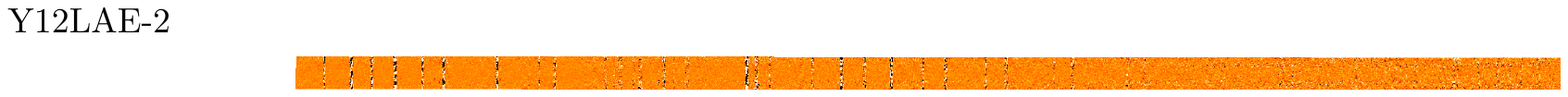}\\
    
    \vspace{10pt}
    \includegraphics[width=2\columnwidth, bb=30 668 552 814]{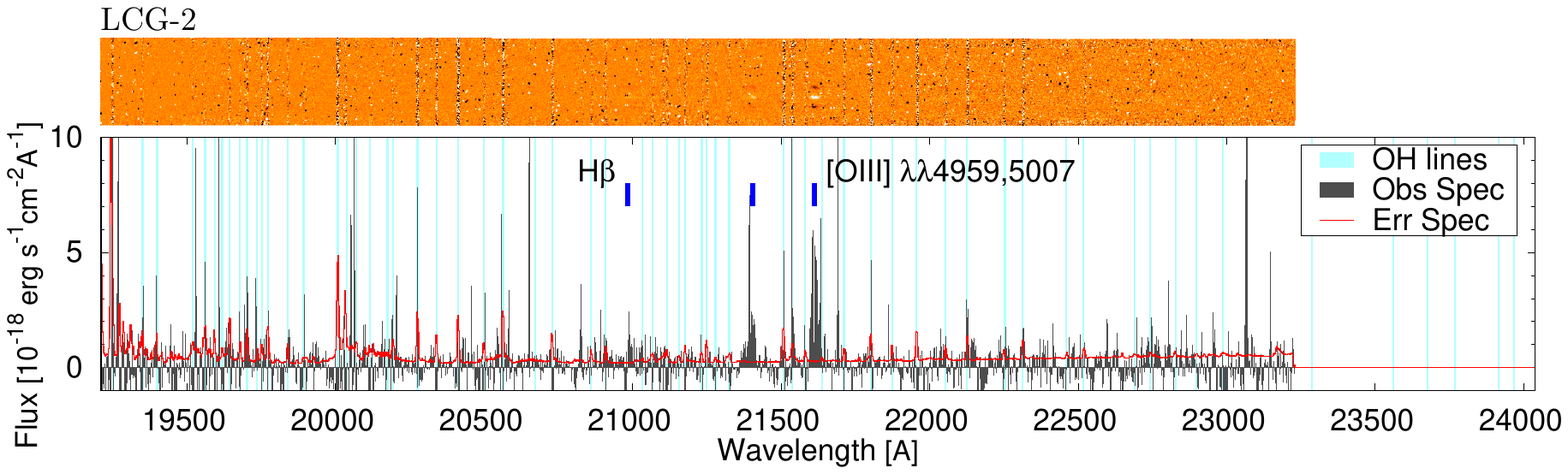}\\
    
    \vspace{10pt}
    \includegraphics[width=2\columnwidth, bb=30 668 552 814]{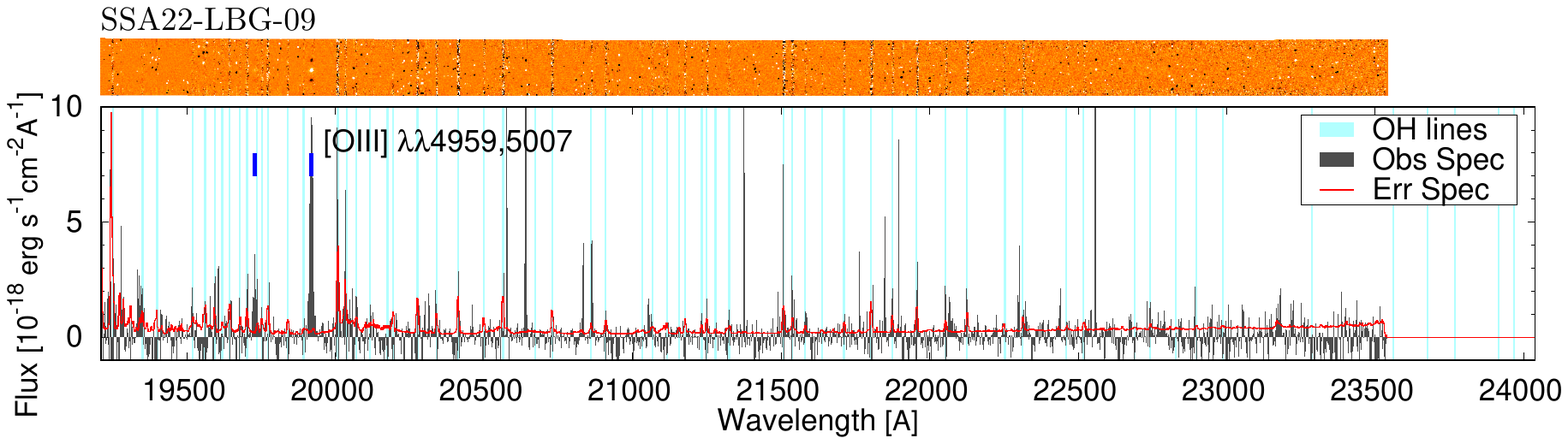}\\
    
    \vspace{10pt}
    \includegraphics[width=2\columnwidth, bb=30 668 552 814]{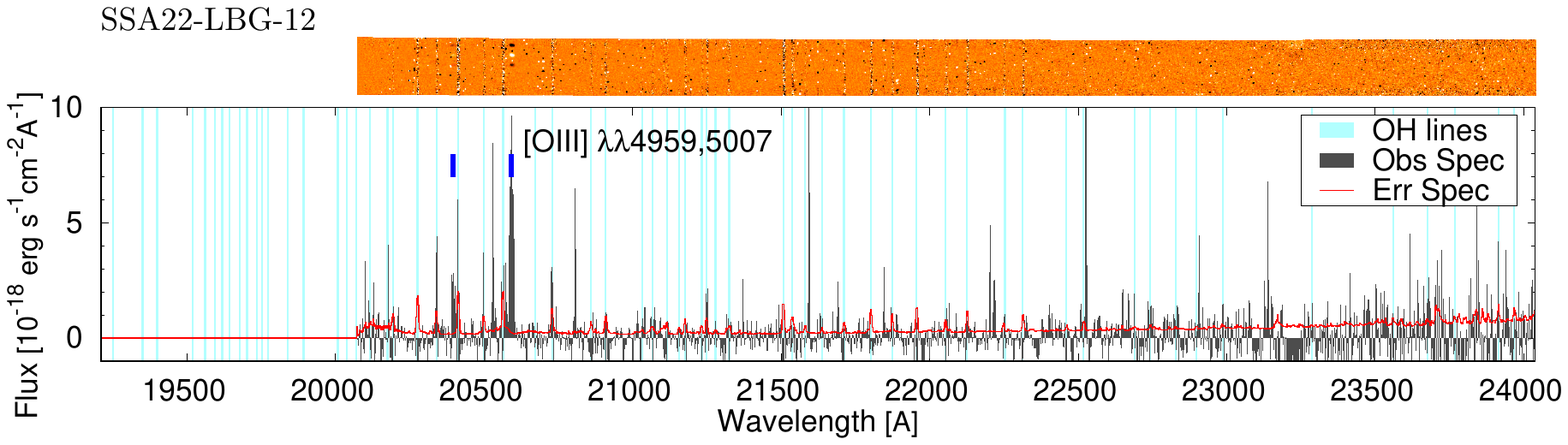}\\
    
    \contcaption{ Reduced 2-D images and 1-D spectra. }
    \label{ap:fig1cont5}
\end{figure*}

\begin{figure*}
    \includegraphics[width=2\columnwidth, bb=30 774 552 814]{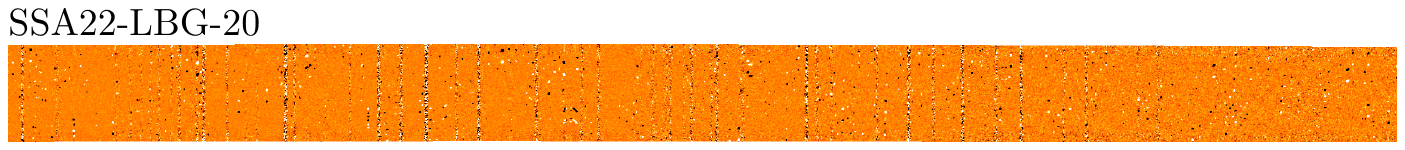}\\
    
    \vspace{10pt}
    \includegraphics[width=2\columnwidth, bb=30 774 552 814]{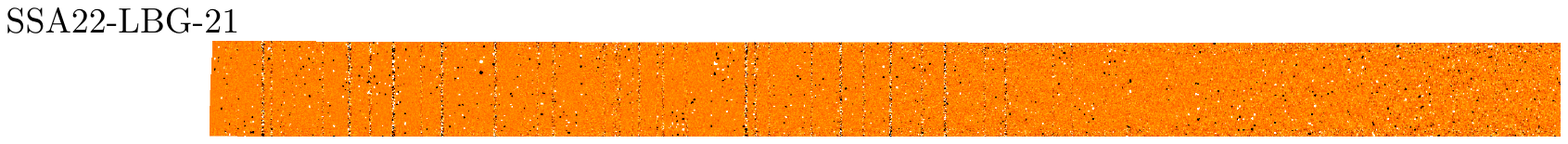}\\
    
    \vspace{10pt}
    \includegraphics[width=2\columnwidth, bb=30 668 552 814]{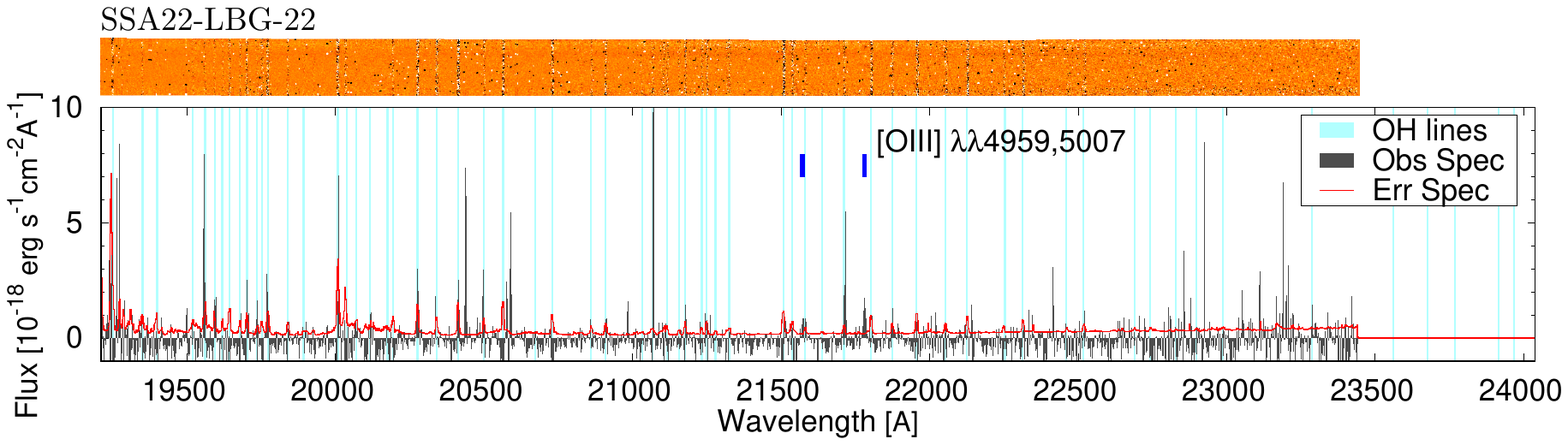}\\
    
    \vspace{10pt}
    \includegraphics[width=2\columnwidth, bb=30 784 552 814]{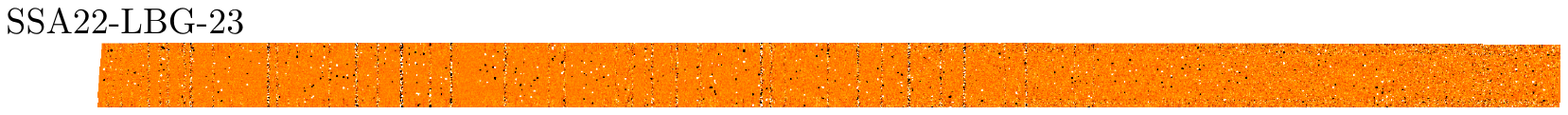}\\
    
    \vspace{10pt}
    \includegraphics[width=2\columnwidth, bb=30 784 552 814]{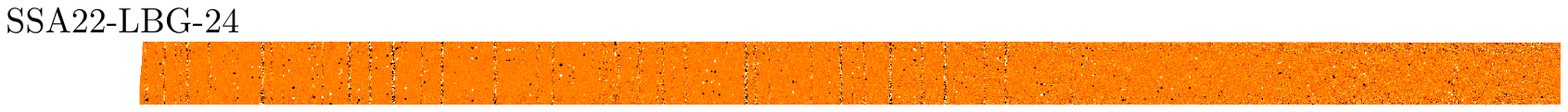}\\
    
    \vspace{10pt}
    \includegraphics[width=2\columnwidth, bb=30 789 552 814]{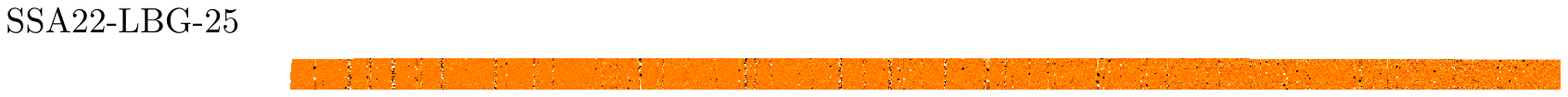}\\
    
    \vspace{10pt}
    \includegraphics[width=2\columnwidth, bb=30 668 552 814]{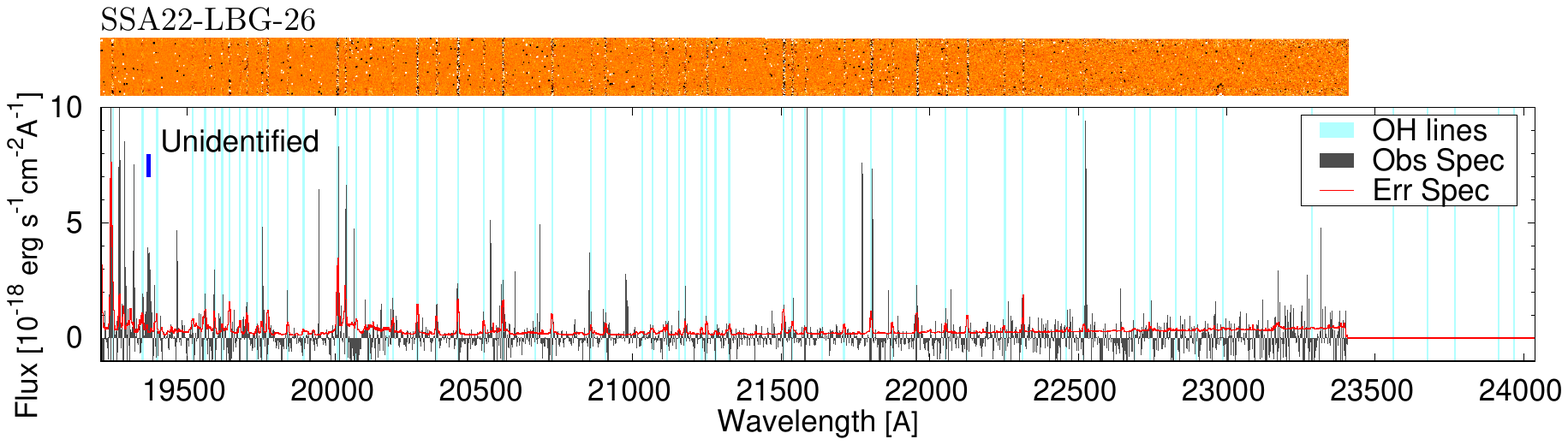}\\
    
    \vspace{10pt}
    \includegraphics[width=2\columnwidth, bb=30 789 552 814]{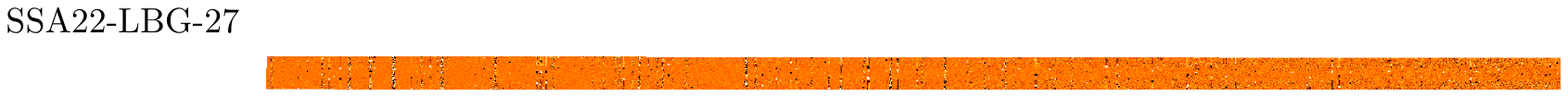}\\
    
    \vspace{10pt}
    \includegraphics[width=2\columnwidth, bb=30 784 552 814]{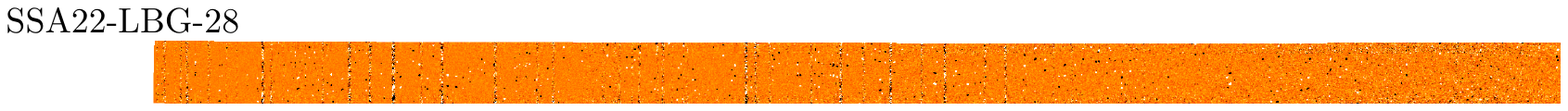}\\
    
    \vspace{10pt}
    \includegraphics[width=2\columnwidth, bb=30 784 552 814]{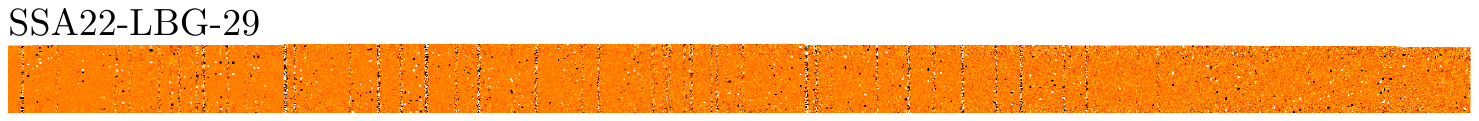}\\
    
    \vspace{10pt}
    \includegraphics[width=2\columnwidth, bb=30 784 552 814]{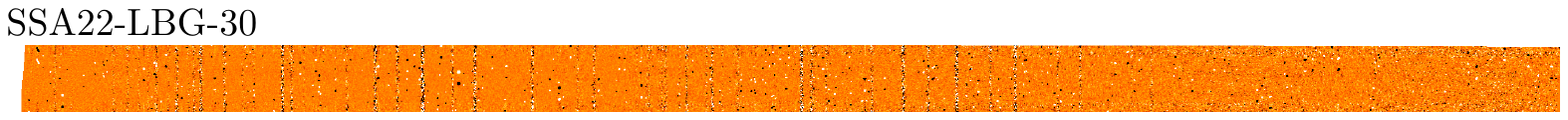}\\
    
    \contcaption{ Reduced 2-D images and 1-D spectra. }
    \label{ap:fig1cont6}
\end{figure*}

\begin{figure*}
    \includegraphics[width=2\columnwidth, bb=30 668 552 814]{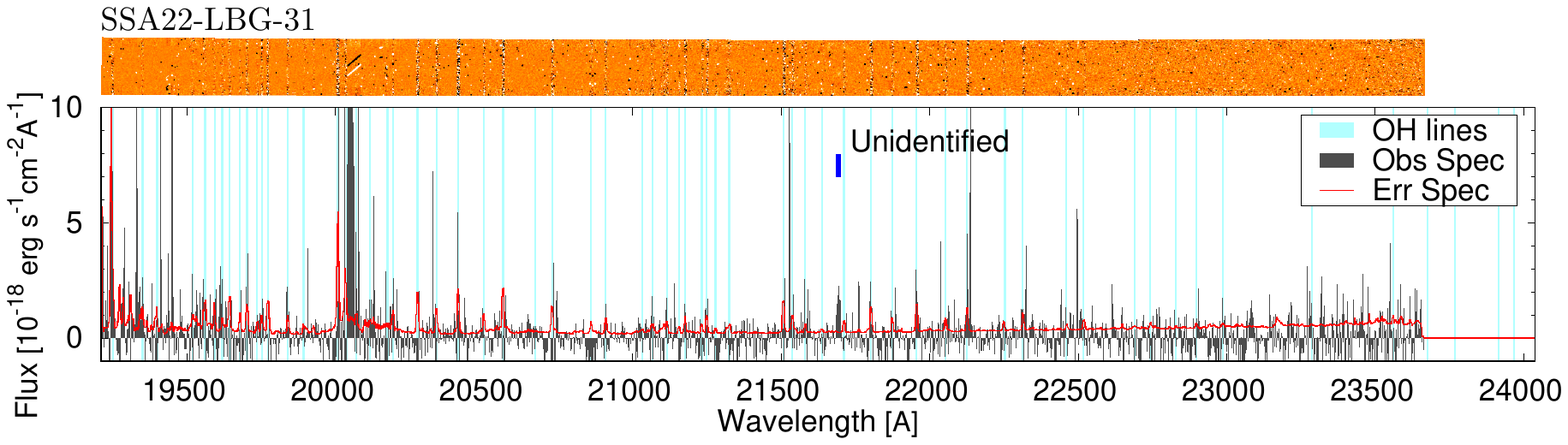}\\
    
    \vspace{10pt}
    \includegraphics[width=2\columnwidth, bb=30 784 552 814]{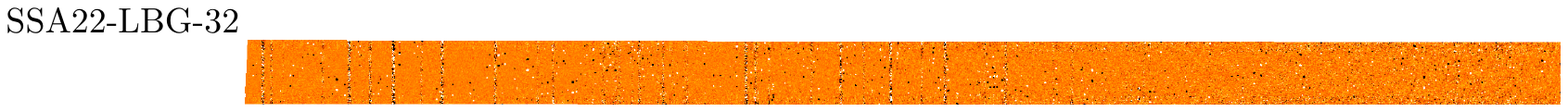}\\
    
    \vspace{10pt}
    \includegraphics[width=2\columnwidth, bb=30 774 552 814]{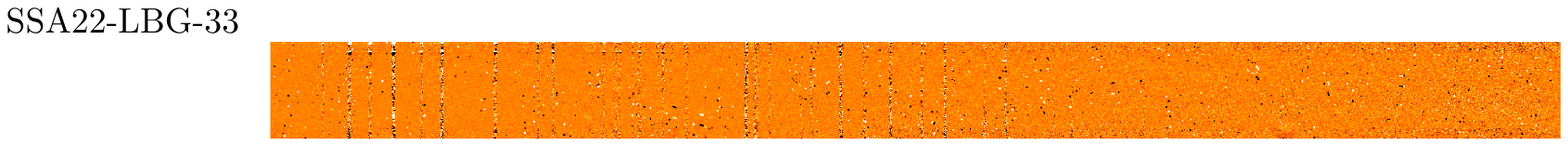}\\
    
    \vspace{10pt}
    \includegraphics[width=2\columnwidth, bb=30 784 552 814]{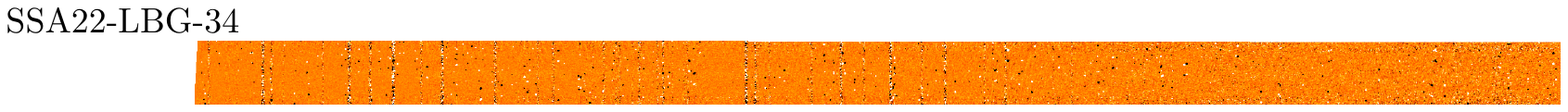}\\
    
    \vspace{10pt}
    \includegraphics[width=2\columnwidth, bb=30 789 552 814]{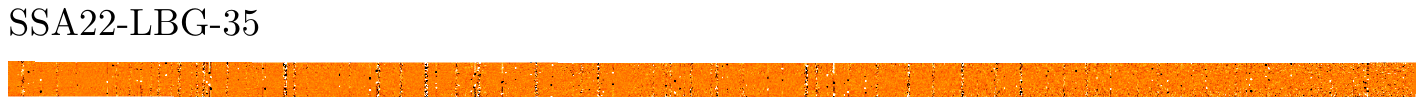}\\
    
    \vspace{10pt}
    \includegraphics[width=2\columnwidth, bb=30 774 552 814]{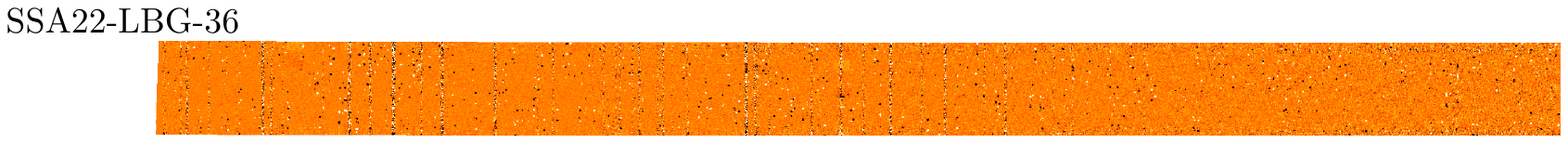}\\
    
    \vspace{10pt}
    \includegraphics[width=2\columnwidth, bb=30 789 552 814]{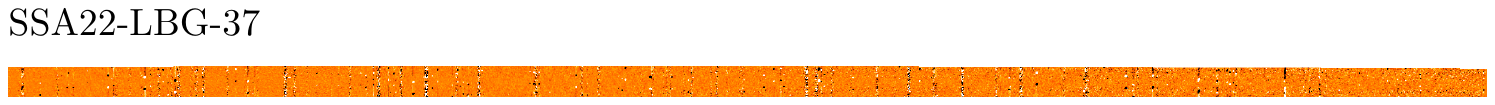}\\
    
    \vspace{10pt}
    \includegraphics[width=2\columnwidth, bb=30 668 552 814]{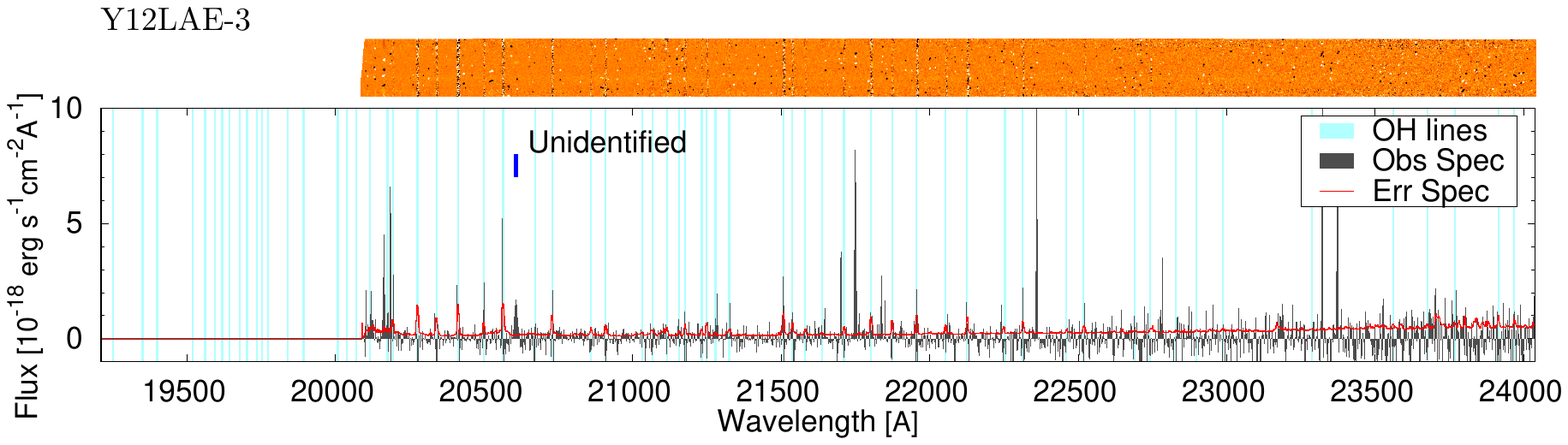}\\
    
    \contcaption{ Reduced 2-D images and 1-D spectra. }
    \label{ap:fig1cont7}
\end{figure*}


\bsp	
\label{lastpage}
\end{document}